\documentstyle[12pt]{article}
\input amssym.def
\font\teneusm=eusm10                    
\font\seveneusm=eusm7                   
\font\fiveeusm=eusm5                    
\newfam\eusmfam
\textfont\eusmfam=\teneusm
\scriptfont\eusmfam=\seveneusm
\scriptscriptfont\eusmfam=\fiveeusm
\def\sh#1{{\fam\eusmfam\relax#1}}
\input amssym
\newcommand{\be}{\begin{equation}}
\newcommand{\ee}{\end{equation}}
\newcommand{\ba}{\begin{eqnarray}}
\newcommand{\ea}{\end{eqnarray}}
\newcommand{\ban}{\begin{eqnarray*}}
\newcommand{\ean}{\end{eqnarray*}}
\newcommand{\brr}{\begin{array}}
\newcommand{\err}{\end{array}}
\newcommand{\bc}{\begin{center}}
\newcommand{\ec}{\end{center}}
\newcommand{\bea}{\begin{eqnarray}}
\newcommand{\eea}{\end{eqnarray}}
\newcommand{\bean}{\begin{eqnarray*}}
\newcommand{\eean}{\end{eqnarray*}}
\newcommand{\sss}{\scriptstyle}

\newcommand{\rref}[1]{(\ref{#1})}
\def\0{\nonumber}
\newcommand{\pai}{{\textstyle{1\over{2\pi i}}}}
\newcommand{\za}{{z_\alpha}}
\newcommand{\zb}{{z_\beta}}

\def\A{{\cal A}}
\def\B{{\cal B}}
\def\C{{\cal C}}

\def\E{{\cal E}}
\def\R{{\cal R}}
\newcommand{\F}{{\cal F}}

\def\L{{\cal L}}
\newcommand{\M}{{\cal M}}

\def\O{{\cal O}}

\def\Q{{\cal Q}}

\def\pg{{\goth p}}
\newcommand{\Pg}{{\goth P}}
\newcommand{\hg}{{\goth h}}

\def\s{{\sigma}}
\def\al{{\alpha}}
\def\o{{\omega}}
\def\cb{{\bf c}}
\def\sb{{\bf s}}
\def\ct{{C_\tau}}
\def\ai{{a_i}}
\def\bi{{b_i}}
\newcommand{\HH}{{\Bbb H}}

\newcommand{\CC}{{\Bbb C}}
\newcommand{\PP}{{\Bbb P}}

\newcommand{\Slc}{SL(2, {\Bbb C}\, )}

\newcommand{\Sltc}{SL(3, {\Bbb C}\, )}

\newcommand{\Slnc}{SL(n, {\Bbb C}\, )}
\newcommand{\slnc}{sl(n, {\Bbb C}\, )}

\newcommand{\del}{\partial}
\newcommand{\delb}{{\bar\partial}}

\begin{document}
\begin{flushright}
SISSA-ISAS 27/93/EP\\
hep-th/9303064
\end{flushright}
\vskip0.5cm
\centerline{\LARGE Liouville and Toda field theories}
\vskip0.2cm
\centerline{\LARGE on Riemann surfaces}
\vskip1.5cm
\centerline{\large  E. Aldrovandi\footnote{Work supported
by the Danish Natural Science Research
Council.}${}^,$\footnote{e--mail: ettore@mi.aau.dk}}
\centerline{Matematisk Institut, \AA rhus Universitet}
\centerline{Ny Munkegade, DK-8000 \AA rhus C, Danmark}
\vskip0.2cm
\centerline{\large  L. Bonora\footnote{e--mail:
bonora@tsmi19.sissa.it}}
\centerline{International School for Advanced Studies (SISSA/ISAS)}
\centerline{Via Beirut 2, 34014 Trieste, Italy}
\centerline{and INFN, Sezione di Trieste.  }
\vskip5cm
\abstract{We study the Liouville theory on a Riemann surface of genus g
by means of their associated
Drinfeld--Sokolov linear systems. We discuss
the cohomological properties of the monodromies of these systems. We identify
the space of solutions of the equations of motion which are single--valued
and local and explicitly represent them in terms of Krichever--Novikov
oscillators. Then we discuss the operator structure of the quantum
theory, in particular we determine the
quantum exchange algebras and find the quantum conditions
for univalence and locality. We show that we can extend the above discussion
to $sl_n$ Toda theories.}
\newpage
\setcounter{footnote}{0}

\section{Introduction}

A link between the Liouville equation and Riemann surfaces was found first
in mathematics as a clue to the uniformization theory of Riemann surfaces.
Much more recently the Liouville equation has appeared in the theoretical
physics literature in connection with string theory, 2D gravity and
conformal field theory (for reviews of various aspects of Liouville
theory in physics, see \cite{S}).

The Liouville action appears in the Polyakov string theory
path integral when, in order to perform the functional integration over
the metrics of a given Riemann surface, one fixes the gauge freedom by
choosing the so--called conformal gauge. What one is supposed to do next,
for off--critical string theory, is to
integrate over the Liouville field (i.e. quantize the Liouville theory) and
the other relevant modes (ghosts, matter), and finally integrate over
the moduli space. It is evident that uniformization theory must play a
very important role here \cite{ZT}. This ambitious program meets with
formidable obstacles, and, in any case, the approach based
on matrix integrals and on topological field theories is certainly more
effective, for the time being. However Polyakov's path integral remains
a basic suggestion and a basic challenge in string theory. Eventually
one should be able to reconcile the different approaches.

Another context in which the Liouville theory appears is the
Coulomb gas representation of conformal field theories \cite{DF}. For it is
well--known that the Coulomb gas is a set of practical recipes
to construct conformal field theories of a certain type, typically
minimal models, which is nothing
but a manifestation of an underlying Liouville \cite{GN},\cite{CT} or
conformal Toda theory. In this case the Liouville field does not play
the role of a metric, as it does
in string theory; the metric is a fixed one on a fixed Riemann surface and
the most frequently studied case is the one of a flat metric in genus 0.
Although general theorems have been formulated and partial results have been
obtained for conformal field theories on Riemann surfaces, almost nothing is
known concerning the approach to conformal field theories in higher genus
by means of the Liouville or Toda theories, which, we recall, provide a
systematic method to calculate correlation functions.

Whatever the context we consider, it is clear that a better knowledge
of Liouville and Toda theories on Riemann surfaces would be most welcome.
With this motivation we set out to study, in this paper, the
quantization of a Liouville theory on a Riemann surface of fixed genus.
By quantization we mean canonical quantization and to avoid misunderstandings
we recall the quantization procedure in genus zero, more precisely on
a cylindrical topology, presented in \cite{LS}, \cite{BBT}, \cite{ABBP},
\cite{BB}.
There the classical phase space was defined as the space of solutions of
the Liouville equation endowed with the canonical Poisson bracket. It was
shown that this phase space can be represented by means of free bosonic
oscillators, and, at this point, it was elementary, at least in principle,
to quantize it by transforming the free bosonic oscillators into free
bosonic creation and annihilation operators. In this construction a
crucial role is played by the appropriate Drinfeld--Sokolov (DS) linear system
\cite{DS}. The quantum
construction on the other hand hinges upon the quantum group symmetry.

In section 2 of this paper we introduce the two (chiral and antichiral)
DS systems appropriate
for a Liouville theory on a generic genus g Riemann surface with the help
of the concept of analytic connection.
In section 3 we set out to study the properties of the solutions
of the DS systems. In this context we found it very helpful to
be able to explicitly
express the connections $\pg$ and $\bar \pg$, which specify the DS
systems, in terms of bases of meromorphic
differentials on the Riemann surface with two punctures. The
properties of these bases (the Krichever--Novikov bases) are summarized
in subsection 3.1. The rest of section 3 is devoted to the monodromy
of the DS systems: it is particularly pertinent to analyze it in
cohomological terms and, in fact, we show that it is an $\Slc$-valued
cocycle.

Next we turn to the study of the solutions of the Liouville equation
which can be obtained via solutions of the DS systems. Our aim is
to identify the solutions which are local and single--valued on a Riemann
surface with two punctures. In section 4 we discuss single--valuedness, which
imposes two set of constraints on $\pg$ and $\bar \pg$: the first set
tells us that the zero modes of the chiral and antichiral DS systems
around the two punctures and around the homology cycles should be the same
(this is a generalization of the genus 0 constraint);
the second set expresses the fact that the monodromies of the DS systems
must be represented by coboundaries (such type of condition is
not needed in genus 0).
In section 5 we introduce a symplectic structure on the space of
DS connections. We are then able to calculate the exchange algebra
and to impose locality. It turns out that the first set of constraints
required for univalence is first class, while the second set is second
class. Finally we introduce the corresponding Dirac brackets.

In section 6 we pass to quantization. We find the quantum exchange algebra
and the quantum analogs of the above two sets of constraints which
ensure univalence and locality. Section 7 is devoted to the generalization
of the previous results to Toda theories. We think the example of the
$sl_3$ Toda theory is enough to convince the reader that everything works
for these theories too, up to minor technical modifications.

We mentioned above the two physical contexts in which the Liouville equation
appears. It may therefore be interesting to compare the two
corresponding types of solutions.
In section 8 we make a detailed comparison between
the solutions obtained via the DS systems and the uniformizing solution
for a compact Riemann surface. We conclude that the latter is not
included among the former.

Finally Appendices
A and B contain detailed developments omitted in the main text,
Appendix C is devoted to analyzing a non--standard family of solutions of the
Liouville equation and, finally, Appendix D
contains a discussion of the conformal properties of the Bloch wave basis.

The main results of our paper are: 1) the identification of the cohomological
properties of the monodromy of the DS system (section 3);
2) the identification of the constraints \rref{condgamma}, \rref{Fgamma},
\rref{barFgamma}; 3) the exchange algebras and the Dirac brackets of section 5;
4) the quantum exchange algebra and quantum coboundary
conditions \rref{qFgamma}
and \rref{qbarFgamma} of section 6;
6) the generalization to the $sl_3$ Toda field theory in section 7.

\section{The Liouville equation on Riemann surfaces}

The basic objects of our analysis will be the solutions of the
Liouville equation. Let $X$ be a fixed compact Riemann surface of
genus $g\geq 2$\footnote{This limitation will be held throughout the
paper in order to avoid lengthy specifications on genus 0 and 1, which
could be done anyway. The genus 0 case, in particular, has been treated
at length elsewhere.}.
Usually ${\goth U}$ = $\{(U_\alpha, z_\alpha)\}$ will denote
a complex atlas on $X$.
Fixing the complex atlas is tantamount to giving the complex structure,
which will be held fixed throughout.

The Liouville equation is
\bea
\partial \bar \partial\varphi =e^{2\varphi}
\label{L}
\eea
where $\partial= {\partial \over { \partial z}}$
and $\bar \partial= {\partial \over { \partial \bar z}}$, $z$ being any
local coordinate. In \rref{L} we drop the chart label due
to its invariance under a change of local coordinates.
For if ${(U_\alpha, z_\alpha)}$ $\to$ ${(U_\beta, z_\beta)}$
with holomorphic coordinate change
$z_\alpha= f_{\alpha\beta}(z_\beta)$, eq.(\ref{L}) will not change
its form if
\be
\varphi_\beta(z_\beta)= \varphi_\alpha(f_{\alpha\beta}(z_\beta))+
\frac {1}{2} \log |f'_{\alpha\beta}(z_\beta)|^2
\ee
This implies in particular that $e^{2\varphi}$ transforms as a
$(1,1)$--form. We can consider it as the K${\rm \ddot a}$hler form of a
metric on $X$ if $e^{2\varphi}$ is regular. This is true in particular if the
solution happens to be the one coming from the uniformization of $X$.

\subsection{The DS linear system}

To find a large class of solutions of the Liouville equation we proceed as
in genus zero \cite{BBT},\cite{ABBP} and write the linear system
associated to it
\bea
\partial \Q &=& (\pg H - E_+)\Q ,\qquad
\delb\pg = 0\label{DS1}\\
\bar \partial \bar \Q &=& - \bar \Q (\bar \pg H -E_-) ,\qquad
\del\bar\pg = 0\label{DS2}
\eea
where once again we drop the chart label (since we will impose the form
of these equations to hold in any coordinate patch), and
\[
H=\left(\matrix{1&0\cr 0&-1\cr}\right),\quad\quad
E_+=\left(\matrix{0&1\cr 0&0\cr}\right),\quad\quad
E_-=\left(\matrix{0&0\cr 1&0\cr}\right)
\]
The solution matrices $\Q$ and $\bar \Q$ have the form
\[
\Q= \left( \matrix {\s_1 &\s_2 \cr 0 & \s_1^{-1}\cr} \right),\quad\quad
\bar\Q= \left( \matrix {\bar \s_1 &0 \cr \bar \s_2 &\bar \s_1^{-1}\cr} \right)
\]
and locally
\bea
&&\s_1(z)= e^{\int^z \pg(w) dw},
\quad\quad \s_2 (z) =-\s_1(z) \int^z\s_1(w)^{-2}dw\label{sol1}\\
&&\bar \s_1(\bar z)= e^{-\int^{\bar z} \bar \pg(\bar w) d\bar w},
\quad\quad \bar \s_2 (\bar z) =\bar \s_1(\bar z) \int^{\bar z}
\bar \s_1(\bar w)^{-2}d\bar w\label{sol2}
\eea
Then, in a local patch,
\be
e^{-\varphi}= \s M\bar \s, \quad\quad \s =(\s_1,\s_2), \quad\quad
\bar \s= \left(\matrix {\bar \s_1 \cr \bar \s_2}
\right)
\label{lsol}
\ee
is a solution of the Liouville equation, for any constant matrix $M$.

What we have done so far is well--defined in any local chart, but now we have
to define the DS system in a global way on $X$. Let us consider (\ref{DS1})
first. We remark that in order
for $e^{2\varphi}$ to be a $(1,1)$ form, $\s_i$ and $\bar \s_i$ must be
tensors of weight $-{1}/{2}$ for $i=1,2$. This implies that the rows of
$\Q$ have weights $(-1/2,0)$ and $(1/2,0)$. Thus our arena will be the
holomorphic
vector bundle $V=K^{-{1\over 2}}\oplus K^{1\over 2}$, where $K^{1\over 2}$ is
a square root of the canonical line bundle $K$. Since there are many possible
choices, we fix one once for all.

The next step will be to define the DS linear system as an analytic
connection on this vector bundle. This is, we believe, the correct and
easiest way to put a differential equation in a global context.
Generally speaking, an analytic
connection in a holomorphic vector bundle $E$ over $X$ is a map
\[
\nabla :\E \longrightarrow {\E}\otimes_{{\O}_X}\Omega^1_X
\]
where ${\O}_X$ is the sheaf of holomorphic functions on $X$,
$\E$ is the sheaf of holomorphic sections of $E$ and $\Omega^1_X$
the sheaf of holomorphic differentials on $X ,$ i.e. the sheaf of
holomorphic sections of $K.$
Analytic connections do not always exist \cite{atiyah}. According to
a theorem of Weil, their existence is equivalent to $E$ being a direct
sum of indecomposable  analytically flat bundles \cite{atiyah,gun2}.
This is certainly not the case for $V\, ,$ since $c_1(K^{1\over 2})=g-1$
and the genus $g$ is supposed to be $\geq 2$. Thus
in such a case an
analytic connection must be more properly defined as a map
\cite{deligne}
\[
\nabla :\E \longrightarrow
{\E}\otimes_{{\O}_X}\Omega^1_X(*Y)
\]
where now $\Omega^1_X(*Y)$ is the sheaf of meromorphic differentials,
holomorphic outside a subset $Y$ of $X.$\footnote{One can consider also
replacing the sheaf $\Omega^1_X(*Y)$ with ${\cal M}^1_X\, ,$
the sheaf of meromorphic 1-differentials on $X.$}
Allowing for poles trivializes the cohomological obstructions to the
existence of (analytic) connections. Therefore on these general
grounds we expect the DS connection to be meromorphic.

After these general remarks let us go into more detail.
We start from our DS connection in a local chart
\[
\nabla^{DS} =\del +\left( \brr{cc} -\pg(z)&1\\0&\pg(z)\err\right)\, dz
\]
and require this form to be maintained in any other local
chart, that is, applying $\nabla$ to a section of $V$ should give a
(meromorphic) section of $V\otimes K.$ This will give conditions on
the coefficients of the connection. First of all, the ``1'' into the
connection matrix has an invariant meaning if we notice that $1$
is a section of
$\mbox{{\sl Hom}}(K^{1\over 2},K^{-{1\over 2}}\otimes K).$
The condition on $\pg$ is the following transformation rule:
\be
\frac{d}{d z_\beta}\log k_{\alpha\beta}^{1/2}=
\pg_\beta (z_\beta ) - \pg_\alpha (z_\alpha (z_\beta ))
\frac{d z_\alpha}{d z_\beta}
\label{p:transf}
\ee
where $k_{\alpha\beta}=d z_\beta /d z_\alpha$ are the transition
functions of $K$ and $k_{\alpha\beta}^{1/2}$ is a suitably chosen
collection of square roots defining $K^{{1\over 2}}\, .$ The condition
\rref{p:transf} has the form of a relation in \v{C}ech cohomology.
Indeed, introducing
$c_{\alpha\beta}=\frac{d}{d\, z_\beta}\log k_{\alpha\beta}^{1/2}\, ,$
we easily have $\{ c_{\alpha\beta}\}\in Z^1({\goth U},\Omega^1_X)$ and
\rref{p:transf} takes the form
$\{ c_{\alpha\beta}\} = \delta \{ \pg_\alpha\}\, ,$ where $\delta$ is the
\v{C}ech coboundary map.
Thus $\{ \pg_\alpha\}\in C^0({\goth U},\Omega^1_X)$ is the cochain whose
coboundary is $\{ c_{\alpha\beta}\}.$ However, it is not difficult to
see that the above relation cannot take place, since the cocycle
$\{c_{\alpha\beta}\}$ is the one defining the Chern class of $K^{{1\over 2}}$
\cite{gun1}, so that \rref{p:transf} can be realized only if we take
$\{ \pg_\alpha\}\in C^0({\goth U},{\cal M}^1_X).$ This confirms our
statement that an analytic connection will in general be meromorphic.
Meromorphic connections on $K$ are treated in \cite{hs}.

Analogous things can be repeated for the antiholomorphic DS (\ref{DS2}).
Once this is done our linear system is well--defined on $X$.

\subsection{The problem}

Therefore from now on our problem is:

{\it
1) to parametrize the space ${\goth F}$ of solutions of the
Liouville equation determined through the recipe (\ref{lsol});

2) to define a symplectic structure in ${\goth F}$ and to identify the subspace
${\goth F}_0\subset {\goth F}$ of solutions which are single--valued and local
with respect to this structure;

3) to quantize ${\goth F}_0$.}

\section{Explicit representation and properties of the DS system}

We set out to solve the problem just formulated.
To this end it is very convenient to rely on an explicit representation of the
connections $\pg$ and $\bar \pg$. We saw that they must be meromorphic
connections on a spin $1/2$ bundle.
In analogy with the genus zero treatment, we will
choose the simplest possible arrangement, that is we will suppose that
all their poles are concentrated only in two generic points $P_+$ and $P_-$ of
$X$. Next we have to parametrize the space of these connections.
Using the property that it is an affine space, we can describe all of them
writing
\[
\pg=\Gamma_0 +p\label{pg}
\]
where $\Gamma_0$ is a fixed reference connection and $p$ is a meromorphic
one--form. As for $\Gamma_0$ we may choose
\be
\Gamma_0= -\del \log h\label{Gamma}
\ee
$h$ being a meromorphic section of $K^{1\over 2}$. If $K^{1\over 2}$ represents
an odd theta characteristic the section $h$ can be chosen holomorphic
\cite{fay}. On the other hand any $p$ can be
represented as
\be
p=\sum_k \,p_k\, \omega^k\label{p}
\ee
where $\{\omega^k\}$ is a complete basis of meromorphic differentials
holomorphic outside the set $Y= P_+\cup P_-$ (see below).
The same we can do for $\bar\pg$.
Therefore our two DS systems, and consequently also the solutions of the
Liouville equation given by the reconstruction formula \rref{lsol},
are parametrized by the moments $p_k$ and $\bar p_k$. This is the
parametrization of the space ${\goth F}$ we anticipated above.

Complete bases on $X$ like $\{\omega^k\}$
do exists and were introduced some time ago by I.M.Krichever and S.P.Novikov.
Due to their importance in our analysis, we devote the next subsection
to recalling their definition and properties. For more detailed information
see \cite{kn1},\cite{kn2},\cite{blmr},\cite{kl}.

Finally a remark concerning the modes $p_k$. In genus 0 they are called
free bosonic oscillators. Here they are the closest thing one can define
to free bosonic oscillators, but they are not truly free. The Poisson
brackets \rref{sympl} below reveal the complicated way these modes interact
with the background geometry of the Riemann surface.

\subsection{KN basis in $X$}

In the following $Q,Q',Q_0,..$ will denote points on $X$, but we will often
stick to the habit of denoting these points with local coordinates
$z,z',z_0,...$.
On $X$ let us consider the two distinguished points $P_+$ and $P_-$  and local
coordinates $z_+$ and $z_-$ around them, such that $z_\pm(P_\pm)=0$. On $X$
we can introduce complete bases of meromorphic tensors which are holomorphic
in $X\setminus Y$. In particular we will need a basis of vector fields $e_n$,
functions $A_n$, 1--differentials $\omega^n$ and quadratic differentials
$\Omega^n$. Here $n$ is integer or half--integer according to whether
$g$ is even or odd. The behaviour near $P_\pm$ is given by
\bean
&&A_n(z_\pm) = a^\pm_n z_\pm^{\pm n-g/2}(1+\O(z_\pm))\\
&&\omega^n(z_\pm) = b^\pm_n z_\pm^{\mp n+g/2-1}(1+\O(z_\pm))dz_\pm\\
&&e_n(z_\pm) =
c^\pm_n z_\pm^{\pm n-3/2g+1}(1+\O(z_\pm))\frac{\del}{\del z_\pm}\\
&&\Omega^n(z_\pm) = d^\pm_n z_\pm^{\mp n+3/2g-2}(1+\O(z_\pm))(dz_\pm )^2
\eean
For $|n|\leq g/2$ the definitions of $A_n$ and $\omega^n$ must be modified,
because of the Weiestrass theorem. We set $A_{g/2}=1$, while for $n= g/2-1,...
,-g/2$ the power of $z_-$ is lowered by one in the above definition of $A_n$.
As for $\omega^n$ and $n= g/2-1,...,-g/2$ the power of $z_-$ must be raised
by 1 in the above definition, while $\omega^{g/2}$ is set equal to the third
kind differential
\be
\omega^{g/2}(z_\pm)= \pm \frac {1}{z_\pm}(1+\O(z_\pm))dz_\pm\label{tkd}
\ee
This differential is chosen to be normalized in such a way that the periods
around any cycle are purely imaginary. This implies that the function
\[
\tau(Q) = Re \oint^Q_{Q_0} \omega^{g/2}
\]
is univalent, for a fixed $Q_0\in X$. The level curves of this function
will be denoted $C_\tau$. They reduce to small circles around $P_\pm$ in the
vicinity of these two points.

The above bases elements are uniquely determined up to numerical constants
due to the Riemann--Roch theorem. So we can set for example
$a_n^+=1$, the $a_n^-$'s being then completely determined. We can do the same
for the $c_n^\pm$'s. As for the remaining constants they are fixed by the
duality relations
\bea
\pai \oint_{C_\tau}dz\, A_n(z)\, \omega^m(z)= \delta_n^m\label{dua1}\\
\pai \oint_{C_\tau}dz\, e_n(z)\, \Omega^m(z)= \delta_n^m\label{dua2}
\eea
The Lie brackets of the basis elements $e_n$ are
\be
\relax [e_n,e_m]= C_{nm}^k\, e_k\label{KN}
\ee
Here and throughout the paper summation over repeated upper and lower indices
is understood, unless otherwise stated. One has
\[
C_{nm}^k= \pai\oint_{C_\tau}[e_n,e_m]\Omega^k\0
\]
Equation \rref{KN} define the KN algebra over $X$. Its central extension
is defined by means of the cocycle
\be
\chi(e_n,e_m)= \frac{1}{24\pi i}\oint_{C_\tau}\tilde \chi(e_n,e_m)\label{chi1}
\ee
the integral is over any simple cycle surrounding $P_+$ in an anticlockwise
way. For any two meromorphic vector fields $f= f(z){\del \over {\del z}}$
and $g= g(z){\del \over {\del z}}$, $\tilde \chi (f,g)$ is given by
\be
\tilde \chi(f,g)= \left({\textstyle{1\over 2}} (f'''g-g'''f)-R(f'g-fg')\right
)
dz_+\label{chi2}
\ee
where $'$ denotes derivative with respect to $z_+$ and $R$ is a Schwarzian
connection. Then the extended KN algebra is defined by
\be
\relax [e_n,e_m]=C_{nm}^k\, e_k +t\,\chi(e_n,e_m),\quad\quad [e_n,t]=0
\label{extKN}
\ee

In the following we will also make use of the relations
\[
dA_n= - \gamma_{nm}\omega^m,\quad\quad\quad \gamma_{nm}=\pai\oint_\ct A_ndA_m
\]
and of the definitions
\[
N_i^n= \pai\oint_\ai \omega^n,\quad\quad\quad M_i^n=\pai\oint_\bi \o^n
\]
where $\{\ai,\bi\}$, $i=1,...,g$ is a basis of homology cycles.
It is easy to prove \cite{bt} the relations
\be
N_i^n\gamma_{nm}=0,\quad\quad M_i^n\gamma_{nm}=0\label{rel*}
\ee
The structure constants $C_{nm}^k$ and $\gamma_{nm}$ vanish outside
a finite band of values of $n+m$ around $n+m=0$.

Finally we will need two remarkable relations proven in \cite{bt}:
\bea
&&N_i^nA_n(Q)=:\A_i=\pai\oint_\ai\o^{g/2}\label{remark1}\\
&&M_i^nA_n(Q)=:\B_i=\pai\oint_\bi\o^{g/2}\label{remark2}
\eea
In other words the LHS's are constants that can be explicitly calculated.

As bases for anti--holomorphic tensors we choose the complex conjugate
of the above bases and will distinguish them from the above ones by means
of a bar: $\bar A_n$, etc.. In particular, due to the choice of
normalization for the third kind differential, we have
\be
\bar \A_i=\A_i,\quad\quad\bar \B_i=\B_i\label{rel**}
\ee

Using the explicit representation of
$\pg$ in terms of \rref{p} and \rref{Gamma},
one can better appreciate what are the solutions of Liouville we are
analyzing: they may in general be very singular at $P_\pm$, but this is
no novelty with respect to genus 0. These are the solutions we need in
order to construct a manageable phase space.

\subsection{Properties of the DS solutions}

This subsection is devoted to analyzing a few general properties of
the solutions of the DS system \rref{DS1}.
As we already explained, the DS connection is a map
\[
\nabla^{DS} :\sh V \longrightarrow
{\sh V}\otimes_{{\sh O}_X}\Omega^1_X(*Y)
\]
where $\sh V$ is the sheaf of holomorphic
section of the chosen vector bundle
$V=K^{-{1\over 2}}\oplus K^{{1\over 2}}.$ With
respect to the covering $\goth U$ it is a collection of
meromorphic differential equations. Thus for any open set $U_\alpha\in
{\goth U}$ we can exhibit a fundamental solution $\Q_\al$ of the
differential equation $\nabla^{DS}\Q_\al=0.$
This requires a choice of the integration constants in \rref{sol1}.
Once this is done, $\Q_\al$ is a local frame for $V$ on
$U_\alpha ,$ but when changing local chart, on $U_\alpha\cap U_\beta$
we will have the gluing law
\[
\Q_\al T^\vee_{\alpha\beta}=\left(\brr{cc}k_{\alpha\beta}^{-1/2}&0\\
0&k_{\alpha\beta}^{1/2}\err\right)\cdot\Q_\beta
\]
where $T^\vee_{\alpha\beta}$ is a constant matrix. This follows from the fact
that $\Q_\al$ and the RHS of the above equation both
solve the differential equation $\nabla^{DS}{\cal Q}=0$ on
$U_\alpha\cap U_\beta$ with respect to $\za .$
The duality symbol ``${}^\vee$'' will be explained later on.

Consistency on the triple intersections the matrices
$\{T^\vee_{\alpha\beta}\}$ implies the cocycle condition:
\[
T^\vee_{\alpha\beta}T^\vee_{\beta\gamma}=T^\vee_{\alpha\gamma}
\]
Changing our choice of the local frames $\{\Q_\al\}$ yields a new
collection $\{\Q_\al^{\sss\prime}\}$ such that
\[
\Q_\al=\Q_\al^{\sss\prime}\,C_\alpha
\]
for appropriate constant matrices $\{C_\alpha\}.$ This in turn
produces a new cocycle $T^{{\sss\prime}\vee}_{\alpha\beta}$ such that
\[
T^{{\sss\prime}\vee}_{\alpha\beta}=
C^{-1}_\alpha T^\vee_{\alpha\beta}C_\beta
\]
and therefore describes the same cohomology class.

Thus we interpret the collection $\{ T^\vee_{\alpha\beta}\}$ as
defining a flat rank $2$ vector bundle or, in other terms, a rank $2$
local system. The local system, in turn, is a representation of the
fundamental group into the relevant structure group. Due
to the singularities of the connection, we must be cautious
about what fundamental group we are talking about. The singularities
of the connection come either from those of $\Gamma_0$ or from
the poles of the KN basis. It is easy to see that the former
can only produce poles, while the latter can produce
in general essential singularities at $P_\pm$ in the solutions
of the DS system.
For this reason we had better remove the points $P_+$ and $P_-$. As
a consequence the fundamental group we consider is the fundamental group
of $X^{\sss\prime}=X\setminus Y.$

What we are going to do next is to build
a collection of local fundamental solution matrices for the DS connection, and
subsequently describe in detail the associated monodromy, i.e. the
local system (this will be done in the next subsection).

Let us discuss first the solution $\s_1$ in \rref{sol1}.
We consider the covering
${\goth U}=(U_\alpha ,z_\alpha )$ of $X'$ and a collection $\{Q_\alpha\}$ of
points of $X,$ one for each open neighbourhood $U_\alpha .$
For every $\alpha$ and $Q\in Q_\alpha$ define
\[
\L_\alpha (Q):=\int_{Q_\alpha}^{Q}\pg_\alpha (z_\alpha )\, dz_\alpha
\]
where the integration goes along any path contained in $U_\alpha$ and
joining $Q_\alpha$ and $Q.$ Thus the local solution \rref{sol1} is
written as
\be
{\s_1}_\alpha (Q)=e^{\L_\alpha (Q)}\, ,\;\;\;\;
{\s_2}_\alpha (Q)=
-e^{\L_\alpha (Q)}\, \int_{Q_\alpha}^{Q} e^{-2\L_\alpha (\za )}\,d\za
\label{sol:za}
\ee
Now, to glue two solutions, consider the following situation. Let
$Q$ belong to $U_\alpha\cap U_\beta$ and consider another
point $Q^{\sss\prime}$ still belonging to the intersection
$U_\alpha\cap U_\beta$ and lying, say, on the path from $Q_\alpha$
to $Q.$ Then we have
\bean
\L_\al (Q)&=&\int_{Q_\alpha}^{Q}\pg_\alpha (z_\alpha )\, dz_\alpha
=
\left( \int_{Q_\alpha}^{Q^{\sss\prime}}+\int_{Q^{\sss\prime}}^Q\right)
\pg_\alpha\, d\za\\
&=&\int_{Q_\alpha}^{Q^{\sss\prime}}\pg_\alpha\, d\za +
\int_{Q^{\sss\prime}}^Q\, \pg_\beta\, d\zb +
\int_{Q^{\sss\prime}}^Q
d\log\left(\frac{d\za}{d\zb}\right)^{1/2}\\
&=&\int_{Q_\alpha}^{Q^{\sss\prime}}\pg_\alpha\, d\za +
\int_{Q^{\sss\prime}}^{Q_\beta}\, \pg_\beta\, d\zb +
\int_{Q_\beta}^Q\, p_\beta\, d\zb +
\left. \log\left(\frac{d\za}{d\zb}\right)^{1/2}\right|^Q_{Q^{\sss\prime}}\\
&=&\L_\beta (Q)+\log\left(\frac{d\za}{d\zb}\right)^{1/2}(Q) +
b_{\alpha\beta}
\eean
with
\[
b_{\alpha\beta}=\int_{Q_\alpha}^{Q^{\sss\prime}}\pg_\alpha\, d\za -
\int_{Q_\beta}^{Q^{\sss\prime}}\, \pg_\beta\, d\zb -
\log\left(\frac{d\za}{d\zb}\right)^{1/2}(Q^{\sss\prime})
\]
So we obtain the transformation rule:
\be
{\s_1}_\alpha (Q)=
c_{\alpha\beta}\left(\frac{d\za}{d\zb}\right)^{1/2}(Q)\,
{\s_1}_\beta (Q)
\label{glue1}
\ee
with $c_{\alpha\beta}=e^{b_{\alpha\beta}}\, .$ The
transformation rule \rref{glue1} is meaningful as
$b_{\alpha\beta}$ (or $c_{\alpha\beta}$) is a number, i.e it does not
depend on the point $Q^{\sss\prime}$ used to calculate it. This is
easily seen simply choosing another point $Q^{\sss\prime\prime}$.
If $Q^{\sss\prime}$ and
$Q^{\sss\prime\prime}$ both lie in $U_\alpha\cap U_\beta$,
by using \rref{p:transf} we obtain
\[
\int_{Q^{\sss\prime\prime}}^{Q^{\sss\prime}}\pg_\alpha\, d\za =
\int_{Q^{\sss\prime\prime}}^{Q^{\sss\prime}}\pg_\beta\, d\zb +
\int_{Q^{\sss\prime\prime}}^{Q^{\sss\prime}}
d\log\left(\frac{d\za}{d\zb}\right)^{1/2}
\]
which proves the assertion.

We have given explicitly the above derivation as a sample of the
calculations we use. From now we will be much more succinct.

On a triple intersection $U_\alpha\cap U_\beta\cap U_\gamma$ it
is easy to verify the cocycle condition:
\[
c_{\alpha\beta}c_{\beta\gamma}=c_{\alpha\beta}
\]
Thus we have constructed a collection $\{c_{\alpha\beta}\}$
with values in $\CC^*$ satisfying the cocycle condition on our
Riemann surface $X'.$

Finally we should discuss what happens if we change the reference
collection from $\{Q_\al\}$ to $\{{Q'}_\al\}.$
It does not present any difficulty to see that
$\{ c_{\alpha\beta}\}$ changes by a coboundary, that is
\[
c_{\alpha\beta}\longrightarrow
c^{\sss\prime}_{\alpha\beta}=a_\alpha\, c_{\alpha\beta} {a_\beta}^{-1}
\]
where the non-zero complex number
$a_\alpha = \exp (\int_{Q_\al}^{Q'_\al}p_\alpha).$

{\it In summary, the differential equation
\[
\del \s_1 =\pg\,\s_1
\]
can be solved on the non compact Riemann surface $X^{\sss\prime}=X\setminus Y.$
The solution involves the 1-cocycle $\{ c_{\alpha\beta}\}$ with values
in $\CC^*,$ and we denote by $C$ the corresponding cohomology
class in $H^1(X^{\sss\prime},\CC^*).$} By the usual correspondence, $C$
is a flat line-bundle over $X'$.
Thus $\s_1$ is actually a
meromorphic section of $C\otimes K^{-{1\over 2}}$
(because of the zeros of $h$).

Let us take up next the construction of
$\s_2.$ We use the same scheme as before, namely we consider the point
$Q$ lying in the intersection $U_\alpha\cap U_\beta$ together with
the paths joining it to the reference points $Q_\alpha$ and
$Q_\beta .$ We find
\be
{\s_2}_\alpha (Q)
=d_{\alpha\beta}\, k_{\alpha\beta}(Q)^{-1/2}\, {\s_1}_\beta (Q)+
c_{\alpha\beta}^{-1}\, k_{\alpha\beta}(Q)^{-1/2}\, {\s_2}_\beta (Q)
\label{glue2}
\ee
with
\[
d_{\alpha\beta}=
-c_{\alpha\beta}\int_{Q_\alpha}^{Q'}e^{-2\L_\al (\za )}d\za
+c_{\alpha\beta}^{-1}\int_{Q_\beta}^{Q'}e^{-2\L_\beta (\zb )}d\zb
\]
where $Q\in U_\alpha\cap U_\beta$.
Again, the number $d_{\alpha\beta}$ does not depend on the point used
to calculate it. Therefore the transformation rule \rref{glue2} is
well--defined on $X'.$ It is apparent that the field $\s_2$ does not
simply transform as a spin $-1/2$ tensor. Although the conformal weight $-1/2$
is preserved, $\s_1$ and $\s_2$ get mixed
upon changing the local chart:
\[
\left(\brr{c}\s_{1_\alpha}\\ \s_{2_\alpha}\err \right)=
k_{\alpha\beta}^{-1/2}
\left(\brr{cc}c_{\alpha\beta}&0\\
d_{\alpha\beta}&c_{\alpha\beta}^{-1}\err\right)\cdot
\left(\brr{c}\s_{1_\beta}\\ \s_{2_\beta}\err \right)
\]
or, passing to the collection $\{ \Q_\al\}$ of fundamental matrices
\[
\Q_\al =\left(\brr{cc}k_{\alpha\beta}^{-1/2}&0\\
0&k_{\alpha\beta}^{1/2}\err\right)\cdot
\Q_\beta\cdot
\left(\brr{cc}c_{\alpha\beta}&d_{\alpha\beta}\\
0&c_{\alpha\beta}^{-1}\err\right)
\]
{}From now on we denote by $T=:\{ T_{\alpha\beta}\}$ the collection of
matrices
\[
T_{\alpha\beta}=
\left(\brr{cc}c_{\alpha\beta}&0\\
d_{\alpha\beta}&c_{\alpha\beta}^{-1}\err\right)
\]
For $T$ the cocycle condition
$T_{\alpha\beta}T_{\beta\gamma}=T_{\alpha\gamma}$ holds.
Indeed, written in terms of $d$, this means:
\[
d_{\alpha\gamma}=c_{\beta\gamma}\, d_{\alpha\beta} +
c^{-1}_{\alpha\beta}\, d_{\beta\gamma}
\]
which can be checked straightforwardly in
the triple intersection $U_\alpha\cap U_\beta\cap U_\gamma .$

It remains for us to examine what happens when we change the reference point
collection $\{Q_\al\}$.
Passing to the reference collection $\{Q_\al^{\sss\prime}\},$
$\{d_{\alpha\beta}\}$ transforms as
\[
d_{\alpha\beta}=a_\alpha^{-1}a_\beta^{-1}d_{\alpha\beta}^{\sss\prime}
+c_{\alpha\beta}^{\sss\prime}f_\alpha a_\beta^{-1}
-{c_{\alpha\beta}^{\sss\prime}}^{-1}f_\beta a_\alpha^{-1}
\]
and this relation can be recast in matrix form
\[
T_{\alpha\beta}=
\left( \brr{cc}a_\alpha&0\\ f_\alpha&a_\alpha^{-1}\err\right)\cdot
T_{\alpha\beta}^{\sss\prime}\cdot
\left( \brr{cc}a_\beta&0\\ f_\beta&a_\beta^{-1}\err\right)^{-1}
\]
thus showing that the cocycle $\{T_{\alpha\beta}\}$ is replaced by a
cohomologous one.

{\it Thus the matrices $\{T_{\alpha\beta}\}$ do form an $\Slc$-valued
cocycle. We denote by $T$ the corresponding cohomology class in
$H^1(X' ,\Slc ).$ $T$ is a rank-$2$ flat $\Slc$
vector bundle on $X' ,$ or, as we already said, a rank-$2$ local
system} \cite{gun1,gun2}.

The solution of the DS system we have just produced is to be
properly interpreted as a section of
$K^{-{1\over 2}}\otimes T$ on $X' .$

For the matrices $\{ T^\vee_{\alpha\beta}\}$ introduced above we have
\[
T^\vee_{\alpha\beta}=
\left(\brr{cc}c_{\alpha\beta}^{-1}&
-d_{\alpha\beta}\\0&c_{\alpha\beta}\err\right)
={}^t T_{\alpha\beta}^{-1}
\]
The notation has been chosen in such a way that $T^\vee$ is indeed the
dual of $T.$

The next step is to describe in more detail the structure of the flat
bundle $T$ so constructed. Due to the fact that the cocycle $T$ is
associated to the multi--valuedness of the local
determinations of the solution, it is natural to refer to $T$ as the
{\it monodromy} of the DS system.

\subsection{A description of the monodromy of the DS system}

In this subsection we want to express our monodromy $T$ in a
way as explicit as possible in terms of the variables $p_k$. We start
with a few preliminary remarks.

The flat bundle $T$, i.e. the monodromy of the DS system, is the
same thing as a representation of the fundamental group $\pi_1(X')$
into $\Slc$ (up to conjugation). This follows from the general fact
that for any connected manifold $M$ and any (Lie) group $G$ we have
\[
H^1(M,\, G)\cong \mbox{\sl Hom}(\pi_1(M),\,G)/G
\]
where the quotient is taken with respect to the action of $G$ on
itself by conjugation \cite{kt,gun1}. Given an element $F$ of
$H^1(M,\, G),$ which is a flat bundle on $M$ we will denote by ${\bf F}$
the corresponding element in the other space and call it ``the
characteristic representation associated with $F$''. We shall exploit
the explicit form of the above isomorphism in order to produce
representatives for the various cohomology classes directly in terms
of suitable line integrals over closed paths.

Considered as a vector bundle our monodromy $T$ has
triangular transition functions, so from that point of view
it is an extension
\[
0\longrightarrow C^{-1}\longrightarrow T\longrightarrow
C\longrightarrow 0
\]
where the flat line-bundle $C$ appears as a quotient. In view of the
isomorphism just described it is a representation taking place in the
(lower) Borel subgroup of $\Slc .$ This fact allows us to separately
analyze the components of the representation.

As for the flat line bundle $C,$ in view of the isomorphism
mentioned above, it is an element of
\[
\mbox{\sl Hom}(H_1(X' ) ,\CC^*)
\]
This follows from the fact that being $\CC^*$ commutative, the
homomorphisms of the fundamental group to $\CC^*$ factor through
$\pi_1(X' )/[\pi_1(X'),\pi_1(X' )]\cong H_1(X').$

We recall that the projection of the fundamental group of the non
compact surface $X'$
onto $H_1(X')$ kills the commutator subgroup,
so that the latter is freely generated by
the symbols $a_1,\ldots ,a_g,\, b_1,\ldots ,\, b_g,\, c_0$, where $c_0$
is a small circle surrounding $P_+$\footnote{We make
the slight abuse of language of denoting with the same letter both an
element of the fundamental group and its image in the first homology
group.}.

Thus the line bundle $C$ is the same thing as a character of the first
homology group and it is determined by its value on the generators
$a_1,\ldots ,a_g,\, b_1,\ldots ,\, b_g,\, c_0$ of the group. Below we
will see that the parametrization provided by the KN basis gives a
formula for the character ${\bf c}$ so determined.
But before doing this, we treat, at the same level of generality,
the off-diagonal term in the monodromy.

We recall that $T$ is the extension of $C$ by $C^{-1}. $ This means
 that the extension class represented by $T$
is an element of the cohomology group
\[
H^1(X' ,{\sh C}^{-2})
\]
which means the first cohomology group of $X'$ with values in the
sheaf ${\sh C}^{-2}$ of  {\em locally
constant} sections  of the flat bundle $C^{-2}=C^{-1}\otimes C^\vee $
\cite{gun2}.

This element in
$H^1(X' ,{\sh C}^{-2})$ is determined in the following way. Recall the
identity :
\[
d_{\alpha\gamma}=c_{\beta\gamma}\, d_{\alpha\beta} +
c^{-1}_{\alpha\beta}\, d_{\beta\gamma}
\]
satisfied by the quantities $\{d_{\alpha\beta}\}$ introduced in
\rref{glue2}.
Following \cite{gun2}, we rewrite it by introducing the 1-cochain
$\{s_{\alpha\beta}\}=\{ c_{\alpha\beta}\, d_{\alpha\beta}\},$ thereby
obtaining the identity
\[
s_{\alpha\gamma}=c^2_{\beta\gamma}\, s_{\alpha\beta} +
s_{\beta\gamma}
\]
This identity is in fact the cocycle condition for the 1-cochain
$s=\{ s_{\alpha\beta}\}$ with values in the locally constant sections of
the flat line bundle $C^{-2}.$
However we should verify that our procedure does define an element
of a cohomology group. In other words, a new cocycle
$\{s_{\alpha\beta}^{\sss\prime}\}$ which differs from the previous one by
a coboundary, should be associated essentially to the same data for
our differential equation. The cochain relation
$ s'-s=\delta (f)$ has the explicit form
\[
s_{\alpha\beta}^{\sss\prime}=s_{\alpha\beta}
+f_\beta - c_{\alpha\beta}^2f_\alpha
\]
Now suppose such a collection $\{f_\alpha\}$ is given.
Dividing by $c_{\alpha\beta}$ and taking into account the explicit
form for $d_{\alpha\beta}$ previously found, we see that shifting by a
coboundary precisely amounts to a change in the integration constants
in the indefinite integrals defining $\{{\s_2}_\alpha\}.$ In other
words, this is the same as changing the initial points in the integral
defining $\{{\s_2}_\alpha\}$ in \rref{sol:za}.

Thus the off-diagonal element of the monodromy also has a cohomological
interpretation.

Let us now give an explicit representation for the classes in
$H^1(X',\CC ^*)$
and $H^1(X', {\sh C}^{-2})$
in terms of the KN parametrization. More precisely, for any
element $\gamma$ of the fundamental group, we provide
representatives ${\bf c}_\gamma$ and
${\bf s}_\gamma$ in terms of line integrals. Let
\[
p=\Gamma_0+p_n\, \omega^n
\]
where $\Gamma_0$ is given by \rref{Gamma}. We recall that the sum
over repeated indices is understood.
Now we plug this expansion in the local expression \rref{sol:za} for
$\s_1.$ We keep the usual definitions and notations for the covering
$\goth U$ and the reference points $\{Q_\alpha\}.$ With an easy
integration we find
\[
{\s_1}_\alpha (Q)=e^{\L_\al (Q)}=
h_\alpha (Q)^{-1}h_\alpha (Q_\al )
\exp\left(p_n \int_{Q_\al}^Q \omega^n\right)
\]
$h_\alpha$ is the determination of $h$ in the chart $U_\alpha .$
Consequently we get
\[
c_{\alpha\beta}=h_\alpha (Q_\al )h_\beta (Q_\beta )^{-1}
\exp\left( p_n \int_{Q_\al}^{Q_\beta} \omega^n\right)
\]
Now we exploit the description of the fundamental group of
$X'$ by means of chains of open sets (see Appendix A).
We fix a base point $Q_0\in X'$ and an open set $U_0$ containing it.
We consider a path $\gamma$ and a covering chain
$(U_{\alpha_0},U_{\alpha_1},\cdots ,U_{\alpha_n},U_{\alpha_0}),$
$U_{\alpha_0}=U_0.$
The character ${\bf c}$ associated with the cocycle $\{ C_{\alpha\beta}\}$
is given by the formula (see Appendix A)
\[
{\bf c}_\gamma =\prod_{i=0}^n c_{\alpha_i\alpha_{i+1}}
\]
which, together with explicit form for the 1-cocycle quoted above
yields
\be
{\bf c}_\gamma=\exp\left(p_n \oint_{\gamma} \omega^n\right)
\label{C}
\ee
Being a character of the first homology group of $X' ,$ ${{\bf c}}$ is
defined by its values on the generators:
\bean
{\bf c}_{a_i}&=&\exp (2\pi i\, N^n_i\,p_n)\\
{\bf c}_{b_i}&=&\exp (2\pi i\, M^n_i\,p_n)
\eean
where the numbers $M_i^n$ and $N_i^n$ where defined in subsection 3.1 and
are in principle
explicitly computable, using the concrete expression for the KN basis
\cite{blmr,kl} in terms of $\theta$-functions and prime form
of $X,$ the compact completion of $X' .$

The value of ${\bf c}$ on the cycle $\ct\equiv c_0$ around
the puncture $P_+$ has a very simple form:
\[
{\bf c}_{c_0}\equiv \cb_0=\exp\left( p_n \oint_{c_0}\omega^n\right)
=e^{2\pi i p_{g/2}}
\]

Next we have to characterize the off-diagonal element of the monodromy
in the same terms as we did for $\cb.$
This is a rather long calculation and it is postponed to the Appendix A.
Here we record the result. The representative ${\bf s}_\gamma$
corresponding to the closed path $\gamma$ is
an algebraic cocycle for the group $\pi_1(X')$ with values in $\CC ,$
that is a map ${\bf s}:\pi_1(X')\rightarrow\CC$ satisfying the
cocycle condition
\[
{\bf s}_{\gamma_1\cdot\gamma_2}=
{\bf c}^2_{\gamma_2}{{\bf s}}_{\gamma_1}+{\bf s}_{\gamma_2}
\]
or, in technical terms, what is called a {\em crossed homomorphism\/}
of $\pi_1(X')$ into $\CC$ with respect to the character
${\bf c}^{-2}.$ The set of all crossed homomorphisms, modulo the
trivial ones, forms a group denoted $H^1(\pi_1(X'), {\bf c}^{-2})$
which is actually isomorphic to $H^1(X' ,{\sh C}^{-2})$ (see Appendix
A and \cite{gun2}). The actual element determined by the DS system is
given by
\be
{{\bf s}}_\gamma = -h_0(Q_0)^{-2}e^{2p_k\oint_\gamma\omega^k}
\oint_\gamma e^{-2p_k \int_{Q_0}^z \omega^k }h(z)^2
\label{sgamma}
\ee
where $\gamma$ is a closed path on $X'$ based  at $Q_0$ covered
by an appropriate chain of open sets. The $\sb_\gamma$ when $\gamma=c_0$
will be denoted for simplicity $\sb_0$.

A few comments are in order.
The integral on the RHS of \rref{sgamma}, as it stands, should be properly
defined on the universal cover of $X' .$ This is due to the fact that
the integrand is multivalued on the surface. Thus formula \rref{sgamma} can
be read in two ways. Interpreting the integral on the RHS as an
integral over the universal covering space, \rref{sgamma} becomes an
instance of the fact that elements of the group
$H^1(\pi_1(M),{\bf F})$ can be represented by means of differential
1-forms on the universal covering space of $M$ ``twisted'' by the
character ${\bf F}$
\cite{gun2}. On the other hand, \rref{sgamma} says that the LHS can be used
to {\em define} the integral on the RHS, thereby giving a full meaning
to a way of naively continuing the integrals in \rref{sol:za}, outside
their domain of definition, along a complete path on the surface.

An immediate formal manipulation of the integral formula \rref{sgamma}
so obtained yields very easily the cocycle condition for
$\sb_\gamma .$ The latter will rigorously follow from the formulas in
Appendix A.

Finally we can do the same for the anti--holomorphic DS system \rref{DS2}.
In particular we find another representation of $\pi_1(X')$, with
representatives of the generators which can be written in the following
form:
\be
\bar{\bf c}_\gamma=\exp\left(-\bar p_n \oint_{\gamma} \bar\omega^n\right)
\label{barC}
\ee
and
\be
\bar{{\bf s}}_\gamma = \bar h_0(Q_0)^{2}e^{-2\bar p_k\oint_\gamma\bar \omega^k}
\oint_\gamma e^{2\bar p_k \int_{Q_0}^{\bar z}\bar \omega^k }h(\bar z)^{-2}
\label{barsgamma}
\ee

\section{Single--valued solutions}

In the previous section we clarified the geometrical meaning and found
explicit expression for the cocycles $\cb_\gamma$ and $\sb_\gamma$
for any cycle $\gamma$.
In this section we will select in the space ${\goth F}$ the solutions
of the Liouville equation which
are single--valued on $X'$. As we will see
this corresponds to putting constraints on ${\goth F}$.
We will proceed in two steps.
First we find the conditions for single--valuedness around $P_+$, then
around the homotopy generators of $X$.

\subsection{Single--valuedness around $P_+$}

We use the reconstruction formula \rref{lsol} and determine the
conditions for the solution to be single--valued around $P_+$.
The problem is the same as in genus 0 and we will simply summarize
the procedure. We start from the solutions
\be
\s(Q)= (\s_1(Q),\s_2(Q)),\quad\quad\quad
\bar \s(Q)= \left(\brr{c} \bar\s_1(Q)\\ \bar\s_2(Q)\err\right)
\ee
of the DS systems \rref{DS1} and \rref{DS2}, respectively.
\bea
&&\s_1(Q)= e^{\int^Q_{Q_0} \pg(z) dz},
\quad\quad \s_2 (Q) =-\s_1(Q) \int^Q_{Q_0}\s_1(z)^{-2}dz\label{sigma12}\\
&&\bar \s_1(Q)= e^{-\int^{Q}_{Q_0} \bar \pg(\bar z) d\bar z},
\quad\quad \bar \s_2 (Q) =\bar \s_1(Q) \int^{Q}_{Q_0}
\bar \s_1(\bar z)^{-2}d\bar z\label{barsigma12}
\eea
where the only difference with respect to \rref{sol1} and \rref{sol2}
is that we have
fixed the initial integration point $Q_0$ once for all: we fix a
curve $\ct$ and we understand that $Q_0$ is a fixed point in it. This
allows us to order the points in $\ct$ in an anticlockwise order with
respect to $P_+$ starting from $Q_0$. Such a choice is very convenient and
does not hinder the generality of our results since we showed in the
previous section that nothing really depends on $Q_0.$\footnote{Changing
the base--point is an isomorphism of the fundamental group.}

The monodromy of these solutions is easily found to be
\be
\s(Q+\ct)= \s(Q)\,{^tT}_0,\quad\quad\quad
T_0=\left( \brr{cc} \cb_0 &0\\
             \cb_0^{-1} \sb_0&\cb_0^{-1}\err\right) \label{monodr0}
\ee
and
\be
\bar \s(Q+\ct)=\bar T_0 \bar\s(Q) ,\quad\quad\quad
\bar T_0=\left( \brr{cc}\bar {\cb}_0 &0\\
                \bar{\cb}_0^{-1} \bar{\sb}_0&\bar {\cb}_0^{-1}\err\right)
\label{barmonodr0}
\ee
where $\cb_0, \sb_0, \bar {\cb}_0, \bar {\sb}_0$ were defined in the previous
section. The notation $T_0$ for the monodromy matrix agrees
with the notation of subsection 3.2 of the matrix $T_{\al\beta}$. In $T_0$
the entries are representatives of the cocycles $c$ and $s$ along the
cycle $\ct$.

Now equation \rref{lsol} gives a single--valued solution around $P_+$
if $M$ is given by
\be
M= g_0\rho\bar\rho \bar g_0^{-1}\label{M}
\ee
where $g_0$ diagonalizes ${^tT}_0$
\be
{}{^tT}_0=g_0D_0g_0^{-1},\quad\quad D_0=\left( \brr{cc}\cb_0 &0\\
                                           0&\cb_0^{-1}\err\right),
\quad\quad g_0= \left(\brr{cc} 1&\frac {\cb_0^{-1} \sb_0}{\cb_0^{-1}-\cb_0}\\
                               0&1\err\right) \label{g0}
\ee
and, similarly
\be
\bar T_0=\bar g_0\bar D_0\bar g_0^{-1},
\quad\quad \bar D_0=\left( \brr{cc}\bar \cb_0 &0\\
                                           0&\bar \cb_0^{-1}\err\right),
\quad\quad \bar g_0= \left(\brr{cc} 1&0\\
\frac {\bar \cb_0^{-1} \bar\sb_0}{\bar \cb_0-\bar\cb_0^{-1}} &1\err\right)
\label{barg0}
\ee
The matrices
\be
\rho= e^{qH}, \quad\quad\quad \bar\rho=e^{-\bar q H}\label{rho}
\ee
where $q$ and $\bar q$ are constants, do not play any role here, and they are
only introduced for later purposes.

The result of this construction is that if we introduce
\be
\psi(Q) = \s(Q) g_0 \rho,\quad\quad\quad \bar \psi(Q)= \bar \rho \bar g_0^{-1}
\bar \s(Q)\label{psi}
\ee
on this new basis, {\it the Bloch wave basis}, the monodromy is diagonal
\[
\psi(Q+\ct) =\psi(Q) D_0,\quad\quad\quad
\bar\psi(Q+\ct)= \bar D_0 \bar \psi(Q)
\]
Thanks to this property the solution of the Liouville equation given by
\be
e^{-\varphi}=\psi\bar\psi\label{lsolpsi}
\ee
is single valued around $P_+$ if $D_0\bar D_0$ is the identity matrix, i.e.
if
\be
\cb_0^{-1}= \bar \cb_0, \quad\quad {\rm or}\quad\quad p_{g\over 2}
= \bar p_{g\over 2}\label{cond0}
\ee
This is the constraint on ${\goth F}$ that guarantees single--valuedness
around $P_+$.

\subsection{Single--valuedness around the remaining loops}

The phase space ${\goth F}$ has to be further restricted if
we want the solution
\rref{lsolpsi} to be single--valued on the whole Riemann surface. A simple
way to find the constraints is as follows. Let $\gamma$ be any homotopically
non--trivial loop for $X$, then
\bean
&&\psi(Q+\gamma)= \psi(Q) \rho^{-1}g_0^{-1}\,{^tT}_\gamma\, g_0\rho\\
&&\bar \psi(Q+\gamma)=  \bar\rho \bar g_0^{-1}\,
\bar T_\gamma\, \bar g_0\bar \rho^{-1}\bar \psi(Q)
\eean
Therefore univalence is guaranteed if
\[
{}{^tT}_\gamma M \bar T_\gamma =M
\]
where $M$ is the same as in eq.\rref{M}. A simple calculation shows that
this implies
\bea
&&\cb_\gamma^{-1} = \bar \cb_\gamma\label{condgamma}\\
&&\F_\gamma \equiv \sb_\gamma- \frac {\cb_0^{-1} \sb_0}{\cb_0^{-1}-\cb_0}
(1-\cb_\gamma^2)=0\label{Fgamma}\\
&&\bar\F_\gamma \equiv \bar\sb_\gamma- \frac {\bar\cb_0^{-1} \bar\sb_0}
{\bar \cb_0^{-1}-\bar \cb_0}
(1-\bar\cb_\gamma^2)=0\label{barFgamma}
\eea
The conditions \rref{Fgamma} and \rref{barFgamma} tell us, from a cohomological
point of view, that the cocycles $\sb_\gamma$ and $\bar\sb_\gamma$,
respectively, are coboundaries (see subsection 3.3).

It is clear that, in order to
guarantee univalence of the solutions \rref{lsolpsi},
we have to impose $2g$ such sets of constraints, one for each generator
in $\pi_1(X).$

The constraints \rref{condgamma} can be written
\be
M_i^n\,p_n= \bar M_i^n\bar p_n,\quad\quad N_i^n\,p_n= \bar N_i^n\bar p_n,
\quad\quad i=1,...,g\label{firstconst}
\ee

As for the remaining constraints,
let us denote them, for practical reasons, $\F_r=0$ and $\bar \F_r=0$,
$r=1,....,2g$.

In conclusion, {\it the phase space of the single--valued solutions is
${\goth F}$ restricted by the $2g+1$ conditions \rref{firstconst} and
by the $4g$ additional constraints $\F_r=0$ and $\bar \F_r=0$,
$r=1,....,2g$.}

To conclude this section, we remark that so far we have talked only about
sufficient conditions for single--valuedness. However it is easy to
convince oneself that the constraints given above are also necessary.
Anyhow this will be clear from Appendix B, where a less
simple--minded derivation of the results of this section is given.
In Appendix C we present a family of solutions of the Liouville equation
disconnected from the ones discussed so far. They will not be included
in our phase space and are unexplored from the quantization point of view.

\section{Exchange algebra and locality}

In this section we define a symplectic structure on the phase space
${\goth F}$, calculate the exchange algebra for the Bloch wave basis
and discuss locality for the solutions of the Liouville equation in $X'$.
Throughout the section we fix a curve $\ct$ and a reference
point $Q_0$ on it.

\subsection{The symplectic structure}

A symplectic structure in ${\goth F}$ can be defined
by means of the Poisson bracket
\be
\{p_n,p_m\}= -\gamma_{nm}\label{sympl}
\ee
If we remember that
\[
p_n= \pai \oint_\ct p(z) A_n(z)dz,\quad\quad\quad \pg(Q) = p(Q)+\Gamma_0(Q)
\]
we find immediately
\[
\{\pg(Q),\pg(Q')\}= -\partial \Delta(Q,Q')
\]
where $\Delta(Q,Q')$ is the $\delta$--function appropriate for 0-- and 1--forms
along $\ct$.

The symplectic structure thus defined is degenerate, for we have
\be
\{p_{g\over 2}, p_n\}=0, \quad\quad \{N_i^mp_m , p_n\}=0,\quad\quad
\{M_i^mp_m,p_n\}=0, \quad\forall n,i\label{degen}
\ee
as a consequence of eq.\rref{rel*}.
We can eliminate the degeneracy by enlarging the phase space with the addition
of the a new variable $q$ such that
\be
\{ q, p_n\}=-A_n(Q_0)\label{qpn}
\ee
We have in particular
\be
\{q, p_{g\over 2}\}=-1,\quad\quad \{q, N_i^mp_m\}=-\A_i,\quad\quad
\{q, M_i^mp_m\}=-\B_i\label{qNM}
\ee
Due to \rref{remark1} and \rref{remark2} and the way we normalized the
third kind differential, we see that the degeneracy is eliminated.
The $q$ introduced here is the one appearing in eq.\rref{rho}.

We have to define the symplectic structure also for the
anti--holomorphic degrees of freedom. So we set
\[
\{\bar p_n,\bar p_m\}= -\bar \gamma_{nm}
\]
Moreover we introduce the multi--conjugate variable
\[
\{\bar q, \bar p_n\}=- \bar A_n(Q_0)
\]

All the remaining Poisson brackets vanish
\[
\{p_n, \bar p_m\}=0,\quad\quad \{q, \bar p_m\}=0,\quad\quad \{\bar q, p_n\}=0,
\quad\quad \forall n,m
\]

In Appendix D we investigate compatibility between the symplectic structure
introduced here and the tensorial properties of the various bases introduced
in section 4. In fact one expects the tensor transformation properties
to be generated by the energy--momentum tensor through the above Poisson
brackets. {\em Only the $\psi$ basis fulfills such a compatibility
requirement.}

\subsection{The classical exchange algebra}

The exchange algebra consists of the Poisson brackets of the components
of the $\s$ basis or the $\psi$ basis among themselves, evaluated at two
different points $Q$ and $Q'$ of $\ct$. In the following it is essential
to keep in mind what we said in subsection 4.1 about the ordering of the
points on $\ct$ with respect to $Q_0$. We will write $Q>Q'$ or $Q<Q'$ according
to whether $Q$ comes after or before $Q'$, if we run $\ct$ starting
from $Q_0$ in an anticlockwise way as seen from $P_+$.

The calculation of the exchange algebra is not very different from the genus 0
case (see, for example \cite{BBT}). Therefore we will avoid a detailed
exposition, but we cannot avoid a few specifications.
For example a simple calculation gives
\[
\{\s_1(Q),\s_1(Q')\}= \gamma_{nm} \int_{Q_0}^Q \omega^n(z)dz
\int_{Q_0}^{Q'}\omega^m(w)dw
\]
Using the properties of the KN bases one finds that
the RHS of this equation can be written as
\[
2\pi i \Big( \epsilon_{Q_0}(Q,Q')-\epsilon_{Q_0}(Q,Q_0)+
\epsilon_{Q_0}(Q',Q_0)\Big)
\]
where we have introduced the symbol
\[
\epsilon_{Q_0}(Q,Q')= \theta_{Q_0}(Q,Q')-\theta_{Q_0}(Q',Q)
\]
and the $\theta$ function is defined as follows for any 1--form $\phi$
\[
\oint_\ct dQ' \phi(Q') \theta_{Q_0} (Q,Q')= \int_{Q_0}^Q dQ' \phi(Q')
\]
To simplify our life we choose henceforth both $Q$ and $Q'$ $>Q_0$, so that
\be
\gamma_{nm} \int_{Q_0}^Q \omega^n(z)dz
\int_{Q_0}^{Q'}\omega^m(w)dw= 2\pi i \,\epsilon_{Q_0}(Q,Q')\label{def}
\ee
{}From now on the calculation is straightforward and one finds -- henceforth
we adopt a simplified notation to avoid the awkward factor $\pai$ in the
formulas: therefore, unless otherwise explicitly stated, $\{\,\,,\,\,\}$
will stand for $\pai\{\,\,,\,\,\}$ --
\bea
&&\{\s_1(Q), \s_1(Q')\}= \epsilon_{Q_0}(Q,Q') \s_1(Q) \s_1(Q')\0\\
&&\{\s_2(Q), \s_2(Q')\}= \epsilon_{Q_0}(Q,Q') \s_2(Q) \s_2(Q')
\label{sigmaex}\\
&&\{\s_1(Q), \s_2(Q')\}= -\epsilon_{Q_0}(Q,Q') \s_1(Q) \s_2(Q')
-4\theta_{Q_0}(Q',Q)  \s_2(Q)\s_1(Q')\0
\eea

We eventually need the exchange algebra in the $\psi$ basis. We have
\[
\psi_1= \s_1 e^q,\quad\quad\quad \psi_2 = \s_2 e^{-q} + \s_1
\frac {\cb_0^{-1} \sb_0}{\cb_0^{-1}-\cb_0} e^{-q}
\]
and the calculation is straightforward on the basis of the previous remarks
and the rules of the previous subsection. One obtains
\bea
&&\{\psi_1(Q), \psi_1(Q')\}= \epsilon_{Q_0}(Q,Q') \psi_1(Q) \psi_1(Q')\0\\
&&\{\psi_2(Q), \psi_2(Q')\}= \epsilon_{Q_0}(Q,Q') \psi_2(Q) \psi_2(Q')
\label{psiex}\\
&&\{\psi_1(Q), \psi_2(Q')\}= -\epsilon_{Q_0}(Q,Q')
\psi_1(Q) \psi_2(Q')-\0\\
&&~~~~~~~~~~~~~~~~~~~~~~~~~~~~~~-4\Big(\frac{\cb_0}{\cb_0^{-1}-\cb_0}+
 \theta_{Q_0}(Q',Q)\Big)  \psi_2(Q)\psi_1(Q')\0
\eea
Similarly for the antichiral half one obtains
\footnote{One may wonder why in the antichiral exchange algebra we do not
use the $\bar \epsilon$ and the $\bar \theta$ symbol, i.e. the
$\epsilon$ and $\theta$ distributions expressed in terms of the antichiral
basis. The reason is that, when limited to $\ct$, $\epsilon$
$(\theta)$ and $\bar \epsilon$ ($\bar \theta$) are different
representations of the same objects.}
\bea
&&\{\bar\psi_1(Q),\bar \psi_1(Q')\}=
-\epsilon_{Q_0}(Q,Q')\bar \psi_1(Q) \bar\psi_1(Q')\0\\
&&\{\bar\psi_2(Q), \bar\psi_2(Q')\}= -
\epsilon_{Q_0}(Q,Q') \bar\psi_2(Q)\bar \psi_2(Q')\label{barpsiex}\\
&&\{\bar\psi_1(Q), \bar\psi_2(Q')\}=
\epsilon_{Q_0}(Q,Q')\bar \psi_1(Q)\bar \psi_2(Q')+\0\\
&&~~~~~~~~~~~~~~~~~~~~~~~~~~~~~~~+4\Big(\frac{\bar\cb_0}
{\bar\cb_0^{-1}-\bar\cb_0}+
 \bar\theta_{Q_0}(Q',Q)\Big) \bar \psi_2(Q)\bar\psi_1(Q')\0
\eea

\subsection{Locality}

By local solutions we mean those for which
\[
\{e^{-\varphi(Q)},e^{-\varphi(Q')}\}=0,\quad\quad Q,Q'\in \ct
\]
With the above exchange algebras in the Bloch wave basis we can easily compute
\bea
&&\{\psi(Q)\bar \psi(Q),\psi(Q')\bar \psi(Q')\} =\0\\
&&=4 \frac{\cb_0\bar \cb_0 -\cb_0^{-1}\bar \cb_0^{-1}}
{(\cb_0^{-1}-\cb_0) (\bar\cb_0^{-1}-\bar \cb_0)}
\Big( \psi_1(Q')\bar\psi_1(Q)\psi_2(Q)\bar \psi_2(Q')
-\psi_1(Q)\bar\psi_1(Q')\psi_2(Q')\bar \psi_2(Q)\Big)\0
\eea
Therefore locality is guaranteed if
\[
\cb_0^{-1}= \bar \cb_0
\]
i.e. the same as \rref{cond0}.

In conclusion, {\it the phase space ${\goth F}_0$ of the single--valued and
local solutions is
${\goth F}$ restricted by the $2g+1$ conditions \rref{firstconst} and
by the $4g$ additional constraints $\F_r=0$ and $\bar \F_r=0$,
$r=1,....,2g$.}

\subsection{Other remarkable Poisson brackets}

One may be interested in the exchange algebra in the covering space of the
Riemann surface, in particular in what happens when we Poisson commute
our solutions after going around $P_+$ a certain number of times.
The answer is particularly simple in the Bloch wave basis. Let us set
\[
\psi^{(n)}(Q)= \psi(Q+n\ct)
\]
Then it is easy to calculate the exchange algebra
\bea
&&\{\psi_1^{(n)}(Q), \psi_1^{(m)}(Q')\}=
\Big( \epsilon_{Q_0}(Q,Q') +(n-m)\Big)\psi_1^{(n)}(Q) \psi_1^{(m)}(Q')\0\\
&&\{\psi_2^{(n)}(Q), \psi_2^{(m)}(Q')\}=
\Big(\epsilon_{Q_0}(Q,Q') +(n-m)\Big)\psi_2^{(n)}(Q) \psi_2^{(m)}(Q')
\label{psinex}\\
&&\{\psi_1^{(n)}(Q), \psi_2^{(m)}(Q')\}= -\Big(\epsilon_{Q_0}(Q,Q')
+(n-m)\Big)\psi_1^{(n)}(Q) \psi_2^{(m)}(Q')-\0\\
&&~~~~~~~~~~~~~~~~~~~~~~~~~~~~~~~~~-4\Big(\frac{\cb_0}{\cb_0^{-1}-\cb_0}+
 \theta_{Q_0}(Q',Q)\Big)  \psi_2^{(m)}(Q)\psi_1^{(n)}(Q')\0
\eea

Another interesting question is the calculation of the Poisson brackets
of the constraints $\F_r$ found above. They are essential in order to
compute the Dirac brackets (see below). To this end we have to compute
the Poisson brackets of the cocycles $\sb_\gamma$ with one another.
{}From the previous experience we know that we have to be able to define
an ordering of the points on the curves over which the cocycles in
question are defined. We proceed as follows. We consider the curve $\ct$
passing through a fixed point $Q_0$ and, keeping $Q_0$ fixed, we continuously
deform $\ct$  so that eventually $\ct$ overlap the curves in question (except
possibly for a set of points of measure zero). One can convince oneself
that this is always possible, and can be done in a definite order,
for example $a_1,...,a_g,b_g,...,b_1, c_0$.
This establishes an ordering according to which
we run one after the other the various curves, starting from $Q_0$ and
following the sense determined by the initial curve $\ct$. Therefore,
given two different curves $\gamma_1$ and $\gamma_2$, we can evaluate
\bea
&&\{\sb_{\gamma_1},\sb_{\gamma_2}\}=\cb_{\gamma_1}\cb_{\gamma_2}
\{\oint_{\gamma_1}\s_1^{-2}(z)dz,
\oint_{\gamma_2}\s_1^{-2}(w)dw\}= \label{s1s2}\\
&&~~~~~~~~~~~~~~=4 \cb_{\gamma_1}\cb_{\gamma_2}\oint_{\gamma_1}dz
\oint_{\gamma_2}dw \s_1^{-2}(z)\s_1^{-2}(w)
\epsilon_{Q_0}(z,w)= \pm 4\,\sb_{\gamma_1}\sb_{\gamma_2}\0
\eea
the + (--) sign  depending on $\gamma_1$ coming after (before) $\gamma_2$
according to the above mentioned ordering.

Other useful Poisson brackets are
\bea
&&\{\rho, \cb_\gamma \}= -\rho \cb_\gamma \chi_\gamma H\label{rhocb}\\
&&\{\rho, \sb_\gamma \}= 2(1-\chi_\gamma) \rho \sb_\gamma H \label{rhosb}
\eea

\subsection{Dirac Brackets}

In order to guarantee univalence of the solutions of the Liouville equation
we had to impose, in section 4, the constraints \rref{condgamma}, \rref{Fgamma}
and \rref{barFgamma}. The first set are first class constraints, for
\be
\{e^{-\varphi(Q)}, \cb_\gamma^{-1}-\bar\cb_\gamma\}=
e^{-\varphi(Q)}(\cb_\gamma^{-1}\chi_\gamma - \bar\cb_\gamma\bar \chi_\gamma)
= e^{-\varphi(Q)}(\cb_\gamma^{-1}-\bar\cb_\gamma)\chi_\gamma\label{firstclass}
\ee
where $\chi_\gamma$ is either $\A_i$ if $\gamma=a_i$, or $\B_i$ if
$\gamma=b_i$, or $1$ if $\gamma=\cb_0$, and similarly for the
barred quantities.
The last equality follows from eq.\rref{rel**}.

The $\F$ and $\bar \F$ constraints, on the contrary, are second class. It is
not hard to compute the Poisson brackets
\bea
&&\{ \F_r,\F_s\} = C_{rs},\quad\quad \0\\
&&\{ \bar\F_r,\bar\F_s\} = \bar C_{rs},\quad\quad \0
\eea
while
\[
\{ \F_r,\bar\F_s\}=0
\]
for all $r,s=1,...,2g$. The matrices $C$ and $\bar C$ are nonsingular.
Therefore, for any two functions $F$ and $G$ on the phase space, we can
define the Dirac brackets
\be
\{F,G\}_*= \{F,G\}- \sum_{a,b}\{F, \F_a\} \C^{-1}_{ab}\{\F_b, G\}\label{Dirac}
\ee
where we use the collective notation $\F_a=(\F_r,\bar \F_r)$, $a=1,...4g$ and
$\C$ is the direct sum of $C$ and $\bar C$.

We have to use these brackets in the restricted phase space ${\goth F}_0$
whenever we require univalence. In particular we should be
careful about the first class constraints \rref{condgamma}. However it is
easy to verify that the $\F_a$ constraints Poisson commute with
$\cb_\gamma^{-1}-\bar \cb_\gamma$, therefore eq.\rref{firstclass} continues
to hold even if we replace the Poisson brackets with the Dirac ones.
Let us also notice that passing to the Dirac brackets preserves the
conformal properties of the Bloch wave basis, see Appendix D.

Finally, a remark concerning locality. Locality is defined with respect
to a fixed contour $\ct$. Therefore, the proof of locality we gave
above is perfectly adequate: locality is referred to the initial Poisson
brackets not to the Dirac brackets just introduced \footnote{Imposing
locality with respect to the Dirac brackets would completely distort the sense
of locality as we have seen that the Dirac brackets understand an
arbitrary deformation of the contour $\ct$. Consequently this kind of
locality would mean Poisson commutativity at two generic points, instead
of commutativity with respect to two points on a fixed $\ct$, i.e. {\it
at fixed Euclidean time}.}.
In other words the
logical order should be the following: first secure locality with respect to
the original Poisson brackets, then impose univalence.

\section{Quantization}

Quantizing the Liouville theory is the necessary step to the ultimate
aim of calculating correlation functions. Here we do not arrive that far but
limit ourselves to the preliminary step of exposing the operator structure
of the quantum theory.
This is not a difficult task, as this problem is quite analogous to
the genus zero case and we can follow the procedure outlined in ref.\cite{BB}.
There, quantization was considered on a lattice as
lattice quantization is particularly adapted to reveal the operator structure.
However the
main result can be immediately translated into a continuum language.

A hint for quantization comes from the substitution
\[
\relax [{~~},{~~}]= i\hbar \{{~~},{~~}\}
\]
which means in particular
\[
\relax [ p_n,p_m]= -i\hbar \gamma_{nm}
\]
for the bosonic oscillators.

One can construct a set of consistent rules that lead in particular
to the following quantum exchange algebra in the $\s$--basis.
\be
\s_1(Q) \s_2(Q')= \s_2(Q') \s_1(Q)R_{12}^\pm,\quad\quad
\brr{ccc} + &{\rm when}& Q>Q'\\ -&{\rm when}&Q<Q'\err\label{qexsigma}
\ee
In this section the labels 1 and 2 appended to $\s$ do not represent the
two components of $\s$ as before, but
\[
\s_1= \s\otimes 1,\quad\quad \s_2= 1\otimes \s
\]
and $R^+_{12}=(R_{21}^-)^{-1}$ is the well--known $sl_2$ quantum R matrix in
the defining representation
\be
R^+_{12}=q^{-{1\over 2}} \left(\brr{cccc}q&0&0&0\\
                                       0&1&x&0\\
                                       0&0&1&0\\
                                       0&0&0&q\err\right)\0
\ee
where $q=\exp (-i\hbar)$ and $x=q-q^{-1}$. An algebra like \rref{qexsigma}
is discussed in \cite{SSW}.

After diagonalizing the quantum monodromy matrix one can then define the
Bloch wave basis in just the same way as we did in the classical case, and
find the corresponding quantum exchange algebra
\be
\psi_1(Q) \psi_2(Q')= \psi_2(Q') \psi_1(Q)\R_{12}^\pm(\cb_0),\quad\quad
\brr{ccc} + &{\rm when}& Q>Q'\\ -&{\rm when}&Q<Q'\err\label{qexpsi}
\ee
where $\R^+_{12}(\cb_0))=(\R_{21}^-(\cb_0))^{-1}$ is the quantum R matrix
appropriate
for the $\psi$--basis
\be
\R^+_{12}(\cb_0)=q^{-{1\over 2}} \left(\brr{cccc}q&0&0&0\\
                                       0&1&-xb_0&0\\
                                       0&xa_0&1-x^2a_0b_0&0\\
                                       0&0&0&q\err\right)\label{qR}
\ee
Here
\[
a_0= \frac{\cb_0}{\cb_0-\cb_0^{-1}},\quad\quad
b_0= \frac{\cb_0^{-1}}{\cb_0-\cb_0^{-1}}
\]
and $\cb_0$ is the quantum version of the cocycle denoted with the same
symbol in the previous sections.

\subsection{Quantum locality and univalence}

Quantum locality of
\[
e^{-\varphi(Q)}= \psi(Q)\bar\psi(Q)
\]
can be easily discussed along the lines of ref.\cite{BB}.
It can be easily seen to depend on the condition
\be
\cb_0\bar \cb_0=1\label{loc}
\ee
If this condition can be imposed, locality is guaranteed. We will see that this
is indeed so. But to understand this and the forthcoming points
we have to know the quantum algebra of the $\cb_\gamma$'s and $\sb_\gamma$'s
among themselves and with the remaining degrees of freedom, in particular
we should define the quantum analogs of the Poisson brackets of subsection
5.4. Following the general formulas of \cite{BB}, this is not difficult.
For any $\gamma =a_1,...,a_g,b_1,...,b_g,c_0$, we have
\be
\rho \cb_\gamma = q^{\chi_\gamma H}\cb_\gamma \rho\label{qrhocb}
\ee
while $\cb_\gamma$ commute with any other operator of the theory. We recall
that $\chi_\gamma$ was defined in connection with eq.(\ref{firstclass}).
As for $\sb_\gamma$ we have
\bea
&&\rho \sb_\gamma =q^{2(\chi_\gamma-1)H} \sb_\gamma \rho\label{qrhosb}\\
&&\sb_{\gamma_1}\sb_{\gamma_2}= q^{\pm4}\sb_{\gamma_2}\sb_{\gamma_1}
\label{sbsb}
\eea
The $\pm$ sign in the last equation has the same meaning as in
eq.(\ref{s1s2}).
It is then easy to verify that
\be
\cb_\gamma \bar \cb_\gamma e^{-\varphi(Q)}=
e^{-\varphi(Q)}\cb_\gamma \bar \cb_\gamma\0
\ee
therefore we are allowed to impose the condition
\be
\cb_\gamma\,\bar\cb_\gamma = 1\label{firstclass'}
\ee
which, in particular, guarantees locality when $\cb_\gamma$ coincides
with $\cb_0$\footnote{Actually in order to impose \rref{firstclass'} one has to
suitably normalize the quantum bases $\psi$ and $\bar \psi$.}.

Next we study the condition of quantum univalence. We can impose
for any $\gamma$ the quantum condition
\[
{}{^tT}_\gamma M \bar T_\gamma =q^{-\chi_\gamma}M
\]
where $M$ is the quantum version of eq.\rref{M}. A simple calculation shows
that
this implies \rref{firstclass'} together with
\bea
&&\sb_\gamma= \frac {\cb_0^{-1} \sb_0}{\cb_0^{-1}-\cb_0}
(1-\cb_\gamma^2)\label{qFgamma}\\
&&\bar\sb_\gamma= \frac {\bar\cb_0^{-1} \bar\sb_0}
{\bar \cb_0^{-1}-\bar \cb_0}
(1-\bar\cb_\gamma^2)\label{qbarFgamma}
\eea
The conditions \rref{Fgamma} and \rref{barFgamma} are the quantum analogs for
the cocycles $\sb_\gamma$ and $\bar\sb_\gamma$ to be coboundaries.

In conclusion, if we impose the conditions \rref{firstclass'}, \rref{qFgamma}
and \rref{qbarFgamma}, $\exp (-\varphi)$ is univalent around $\gamma$
up to the numerical factor $q^{-\chi_\gamma}$.

In order to impose the above constraints on the states of the
theory, an approach \'a la BRST should be convenient.

\section{Higher rank Toda field theories}

So far we have implemented a method for constructing
solutions to the Liouville equation based on the DS construction,
in much the same fashion as in \cite{BBT}. There the construction was
for Toda field theories (ToFT) based on an arbitrary
simple Lie algebra ${\goth g}$ in a representation independent way.
Therefore one may wonder if the setting we are considering here
extends to these more general cases. Although we have not yet
worked out the problem in its full generality, we can show (through a
simple example) that the $\Slnc$--ToFT show the same features
described above for the Liouville case. In particular second-class
constraints corresponding to the off-diagonal monodromies will show
up.

To begin with, let us write down the {\em rank n\/} chiral DS system
in the form\footnote{We shall stick to the case $G=\Slnc$ in the
fundamental representation, so that this ``$n$'' is the same as the
rank of the DS linear system.}
\be
\del\Q = (\Pg - \E_+)\Q\label{DSn}
\ee
where $\Pg$ takes its values in the Cartan subalgebra $\hg$ of
$\slnc$ -- i.e. it is a diagonal matrix with vanishing trace -- and $\E_+$
is the constant matrix $\E_+=\sum_{i=1}^{n-1}E_{i\,i+1},$ $E_{ij}$
being the matrix with the $(i,j)$--th entry equal to $1$ and zero
elsewhere. For simplicity we do not explicitly write down here the
corresponding formulas for the anti-chiral part.

Relying on \cite{BBT} and our previous experience, we promote the
collection of DS linear systems defined for each local chart to an
analytic connection $\nabla^{DS}$ on the vector bundle
\[
W=V^{\otimes_S^n}=K^{-\frac{n}{2}}\oplus\cdots\oplus K^{\frac{n}{2}}
\]
where $ \otimes_S^n$ denotes the n-th symmetric tensor power.
Therefore all the remarks  previously made concerning the poles of the
connection conserve their validity here, since $W$ is a direct sum of
line bundles with non vanishing first Chern class. We remark also
that, the set of all DS connections being obviously affine over the
vector space $\hg\otimes\M^1_X,$ we can still introduce the KN
parametrization
\[
\Pg = \Gamma_0 + \sum_k p_k\omega^k
\]
where now the $\{ p_k\}$ are $\hg$--valued modes and $\{\omega^k\}$ is
the usual KN basis of 1--forms.
As in the Liouville case, from the first row of the fundamental
solution $\Q$ we obtain the chiral multiplet $\{\sigma_\al (z_\al )\}$
behaving like a meromorphic section of $W\otimes T$ over
$X'=X\setminus\{P_+,P_-\},$ and $T$ is the appropriate monodromy to be
calculated as in $\Slc $ case.

Let us give some explicit formulas in the relatively simple case
$G=\Sltc .$ The DS connection has therefore the following form
\[
\nabla^{DS}=\del +
\left(\brr{ccc}-\pg_1& 1    &0\\
                 0   &-\pg_2&1\\
                 0   & 0    &-\pg_3\err\right)\, dz
\]
with $\pg_1+\pg_2+\pg_3=0.$ Requiring $\nabla^{DS}$ to define a
connection on $W=K^{-1}\oplus\CC\oplus K$ implies that
\begin{itemize}
\item $-\pg_1$ is a (meromorphic) connection on $K^{-1}=T ;$
\item $-\pg_2$ is a (meromorphic) 1--form;
\item $-\pg_3$ is a (meromorphic) connection on $K.$
\end{itemize}
At this point one could also easily write down explicit local formulas
for $\sigma (z)=(\sigma_1 (z),\,\sigma_2 (z)\,\sigma_3 (z))$ in a
similar form to \rref{sol1} and \rref{sol2} (except that the level of
nested integrals is augmented by one), but the resulting expressions
are not very interesting here. Instead we shall be concerned with the
resulting local monodromy matrices which we write in the form
\be
T_{\alpha\beta}=
\left(\brr{ccc} {c_1}_{\alpha\beta}&   0               &0\\
                {d_1}_{\alpha\beta}&{c_2}_{\alpha\beta}&0\\
                {d_3}_{\alpha\beta}&{d_2}_{\alpha\beta}&{c_3}_{\alpha\beta}
\err\right),\qquad
{c_1}_{\alpha\beta}{c_2}_{\alpha\beta}{c_3}_{\alpha\beta}=1
\label{t3}
\ee
One can easily verify that the cocycle condition for $\{T_{\alpha\beta}\}$
indeed holds, so that it defines a flat $\Sltc$--bundle over $X'.$
Accordingly, the extension it represents is given by the classes
\[
[ s_2 ] \in  H^1(X',{\sh C}_3\otimes{\sh C}_2^\vee ),
\qquad {s_2}_{\alpha\beta}={c_3}_{\alpha\beta}^{-1}{d_2}_{\alpha\beta}
\]
\[
[ s_1 ] \in H^1(X',{\sh C}_2\otimes{\sh C}_1^\vee ),
\qquad {s_1}_{\alpha\beta}={c_2}_{\alpha\beta}^{-1}{d_1}_{\alpha\beta}
\]
whereas $\{ {d_3}_{\alpha\beta}\}$ is such that
\[
[s_3,s_2] \in H^1(X',{\sh C}_3\otimes{\sh F}_1^\vee ),
\qquad  {s_3}_{\alpha\beta}={c_3}_{\alpha\beta}^{-1}{d_3}_{\alpha\beta}
\]
if $F_1$ is the flat bundle whose representative cocycle is given by
\[
{F_1}_{\alpha\beta}=
\left(\brr{cc} {c_1}_{\alpha\beta}&0\\
               {d_1}_{\alpha\beta}&{c_2}_{\alpha\beta}\\
\err\right)
\]
and ${\sh F}_1$ is the corresponding sheaf of locally constant sections, so
that $T$ is thought as the extension
\[
0\longrightarrow C_3\longrightarrow T\longrightarrow F_1
\longrightarrow 0
\]
On the other hand, if $F_2$ is the flat bundle defined by the cocycle
\[
{F_2}_{\alpha\beta}=
\left(\brr{cc} {c_2}_{\alpha\beta}&0\\
               {d_2}_{\alpha\beta}&{c_3}_{\alpha\beta}\\
\err\right)
\]
then
\[
\left[ {F_2}_{\alpha\beta}^{-1}
\left(\brr{c}{d_1}_{\alpha\beta}\\{d_3}_{\alpha\beta}\err\right)\right]
\in
H^1(X',{\sh F}_2\otimes{\sh C}_1^\vee )
\]
and $T$ appears as
\[
0\longrightarrow F_2\longrightarrow T\longrightarrow C_1
\longrightarrow 0
\]
Thus the flat structure coming out from the DS differential equation is an
agreement with the general theory \cite{gun2}, as expected.

Now, using also the antichiral half of the theory, we require the bilinear
form
\[
\s_\alpha (z_\alpha )M_\alpha \bar\s_\alpha (\bar z_\alpha )
\]
to be univalent (up to the tensor transformation properties, of course)
upon changing local chart.
Here $\{M_\alpha \}$ is a collection of constant $\Sltc$
matrices. Quite obviously there is also the representation space picture, so
that the univalence condition translates into the familiar one
\[
M\, {\bar {\bf T}}_\gamma ={\bf T}^\vee_\gamma\, M
\]
for any $\gamma \in \pi_1(X').$ It is clear that identical formulas must
hold in the more general $\Slnc$ case.

If $n=3$ things are still sufficiently simple that one can work out a
componentwise calculation. The result is the following set of necessary and
(obviously) sufficient conditions
\ba
\bar{\cb_i}_\gamma &=& {{\cb_i}_\gamma}^{-1},\quad i=1,2,3\0\\
{\sb_1}_\gamma &=& \frac {\Delta_{21}}{\Delta_{11}}
(1-{\cb_1}_\gamma{\cb_2}_\gamma^{-1})\0\\
\bar{\sb_1}_\gamma &=& \frac {\Delta_{12}}{\Delta_{11}}
(1-\bar{\cb_1}_\gamma\bar{\cb_2}_\gamma^{-1})\0\\
{\sb_2}_\gamma &=& \frac {m_{23}}{m_{33}}
(1-{\cb_2}_\gamma{\cb_3}_\gamma^{-1})\0\\
\bar{\sb_2}_\gamma &=& \frac {m_{32}}{m_{33}}
(1-\bar{\cb_2}_\gamma\bar{\cb_3}_\gamma^{-1})\0\\
{\sb_3}_\gamma &=& \frac {m_{13}}{m_{33}}
(1-{\cb_1}_\gamma{\cb_3}_\gamma^{-1})-
\frac {m_{23}}{m_{33}}{\cb_2}_\gamma{\cb_3}_\gamma^{-1}{\sb_1}_\gamma\0\\
\bar{\sb_3}_\gamma &=& \frac {m_{31}}{m_{33}}
(1-\bar{\cb_1}_\gamma\bar{\cb_3}_\gamma^{-1})-
\frac {m_{32}}{m_{33}}\bar{\cb_2}_\gamma\bar{\cb_3}_\gamma^{-1}
\bar{\sb_1}_\gamma\label{cond3}
\ea
where $m_{ij}$ are the elements of $M$ and $\Delta_{ij}$ is the
subdeterminant relative to $m_{ij}.$ A little thought shows
that these are precisely the conditions for $T$ and $\bar T$ to be both
equivalent diagonal bundles (i.e. direct sums).

Since from \cite{BBT} it follows that the monodromies around $P_+$ are
diagonalizable, this gives values to the ratios appearing in the conditions
\rref{cond3}, which can in turn be taken as constraints with respect to the
remaining generators of $\pi_1(X').$ As a further side--effect, the values
of the ratios in \rref{cond3} force $M$ to be of the form
\be
M=N_+\,D\, N_-
\label{gauss}
\ee
where $N_\pm$ are completely determined upper and lower unipotent matrices.
Thus the resulting indeterminacy is on the diagonal factor $D.$

We notice that this result relies ultimately on the
assumption that $M$ admits a Gauss factorization. Indeed the conditions
\rref{cond3} are meaningful only if $m_{33}\neq 0$ and $\Delta_{11}\neq 0$
and this is precisely
the condition $M$ must satisfy to admit a Gauss factorization in the form
\rref{gauss}. As such, our result must hold in any rank $n,$ for, if
$M\in\Slnc$ exists and can be represented in the form \rref{gauss}, then the
equation
\[
M\, {\bar {\bf T}}_\gamma ={\bf T}^\vee_\gamma\, M
\, ,\qquad \gamma\in\pi_1(X')
\]
can be recast into the form
\[
N_-\, {\bar {\bf T}}_\gamma\, N_-^{-1}=
D^{-1}\, (N_+\, {\bf T}^\vee_\gamma\, N_+^{-1})\, D
\]
which clearly can hold only if both
$N_-\, {\bar {\bf T}}_\gamma\, N_-^{-1}$
and $N_+\, {\bf T}^\vee_\gamma\, N_+^{-1}$ are diagonal for any
$\gamma\in\pi_1(X').$ Conversely, diagonalizing the monodromy around $P_+$
and imposing the constraints for $\Slnc$ clearly produces an intertwiner
admitting a Gauss factorization. Thus our result can be stated more
precisely by saying that diagonalizability of the monodromy is a necessary
and sufficient condition for a solution of the Toda Field Equations in
terms of a Bloch wave basis to exist.

Finally we can carry out the quantization of this Toda theory
in the same way and with the same limitations as for the Liouville
theory in the previous section. However here we omit explicit formulas
which can be easily inferred from those of ref.\cite{bonbo}.

\section{Liouville equation and uniformization}

The emergence of the Liouville equation in the context of uniformization
theory is such a celebrated result in mathematics that we can hardly avoid
clarifying what is similar and what is different in the DS and the
uniformization approaches. To such purpose we devote this last section.

Let us briefly review the uniformization theory of curves
based on pseudogroups and differential equations \cite{gun1,gun5}. Although
it does not yield such strong results as the one based on discontinuous
actions of Fuchsian groups \cite{bers,fk}, in recent times it has become
popular among physicists because it is based on the idea of
symmetry. Remember that we are sticking to the genus $g>1$ case.

By uniformizing we mean finding a collection of functions $\{A_\alpha\}$,
subordinate to the atlas $\{U_\alpha,\za\}$
\[
A_\alpha\,\,:\,\,U_\alpha\longrightarrow\,\, V_\alpha\subset{\PP}^1
\]
which are required to be local homeomorphisms and such that:
\be
A_\alpha(\za)=\frac{a_{\alpha\beta}A_\beta(\zb)+b_{\alpha\beta}}
{c_{\alpha\beta}A_\beta(\zb)+d_{\alpha\beta}}
\label{projtrans}
\ee
In this way, the new complex atlas $\{U_\alpha,A_\alpha\circ\za\}$ is such
that all local charts are connected through projective transformations.
The collection $\{A_\alpha\}$ can be thought of
as a section of a flat $P\Slc$ bundle on $X.$
These remarks are used in \cite{gun1} to
explicitly construct the uniformizing atlas in terms of sections of
adequate vector bundles. The result is as follows. A projective structure
(subordinate to the complex one on $X$) is constructed taking a section
$\{\xi_\alpha=(\xi_{1\alpha},\xi_{2\alpha})\}$ of the bundle
\[
T\otimes K^{-{1\over 2}}
\]
where $K^{{1\over 2}}$ is a square root of
the canonical bundle and $T$ is a flat
$\Slc$ bundle, that is an element $T\in H^1(X,\Slc )$. The coordinate
sections $\{A_\alpha\}$ are
\be
A_\alpha(\za)=\frac{\xi_{1\alpha}(\za)}{\xi_{2\alpha}(\za)}\0
\ee
and $\xi_{1\alpha}\,,\,\xi_{2\alpha}$ are two independent solutions of
the differential equation
\be
\frac{d^2}{d\za^2}\xi_{\alpha}+
\frac{1}{2}u_\alpha(\za)\xi_{\alpha}(\za)=0\, ,\qquad i=1,2
\label{diffeq}
\ee
with normalized Wronskian.
The equation in \rref{diffeq} glues coherently on $X$ as $\{u_\alpha(\za)\}$
is a projective connection \cite{gun1,hs1}, that is a 1-coboundary for the
1-cocycle $\{f_{\alpha\beta},\zb\}d\zb^2$, namely:
\[
\{f_{\alpha\beta},\zb\}d\zb^2=u_\beta(\zb)d\zb^2-u_\alpha(\za)d\za^2
\]
where $\{f_{\alpha\beta},\zb\}$ is the Schwarzian derivative and
$\za =f_{\alpha\beta}(\zb ).$
By Serre duality, $H^1(X,K^2)=0$, so that projective connections certainly
exist and are in one-to-one correspondence with projective structures
\cite{gun1}.

Let us now define the collection of matrices
\be
F_\al (\za)= \left( \brr{cc} \xi'_{1\al}(\za) & \xi_{1\al}(\za)\\
\xi'_{2\al}(\za) & \xi_{2\al}(\za)\err\right)
\ee
One can prove that there exists a flat
vector bundle which supports a truly analytic connection
\be
\nabla_\al= \partial_\al +\Lambda_\al,\quad\quad   \Lambda_\alpha=\left(
\begin{array}{cc}
0 & 1 \\
-\frac{1}{2}u_\alpha(\za) & 0
\end{array}\right)\label{lambda}
\ee
such that (\ref{diffeq}) can be rewritten as a linear system
\[
\nabla_\al F_\alpha=0\0
\]
and $T$ is realized as the holonomy of $\{\nabla_\al\}.$

Once a uniformization of $X$ is found, one would write down a solution
of the Liouville equation \rref{L} by using the projective charts to
pull back on $X$ the standard Poincar\'e metric on the upper half
plane $\HH ,$ obtaining the very classical formula
\be
e^{2\varphi_\alpha}=
\frac{|\partial A_\alpha|^2}{(\mbox{Im} A_\alpha(\za))^2}
\label{class}
\ee
One can verify that \rref{class} defines a $(1,1)$--form by using the
projective transformation \rref{projtrans}.
This requires the structure group to be reduced from $P\Slc$ to
$PSU(1,1)$ and the coordinate sections $\{ A_\al\}$ to take their
values in $\HH .$ We must notice that this is too optimistic, in
general. What one {\em does\/} obtain, is the so--called
``developing map'' \cite{gun10,gold}
\[
f:\tilde X\longrightarrow \Omega\subset\PP^1
\]
where $\tilde X$ is the universal cover of $X,$ which is equivariant
with respect to the action of $\pi_1(X)$ on $\tilde X$ and of a
certain group $\Gamma\subset P\Slc$ on $\Omega .$ We shall not dwell
on the definition of $f$ any longer, except to mention that $f$ is the
global equivariant map corresponding to the section of the flat
$P\Slc$-bundle over $X,$ therefore it is locally a projective chart
\cite{gold}. The homomorphism $\rho :\pi_1(X)\rightarrow \Gamma$ with
respect to which $f$ is equivariant is the holonomy of the projective
structure \cite{gold}.

Since $X$ has genus $g>1,$ $f$ is a covering map, but $\Omega$ is not
analytically equivalent to $\HH$ in general \cite{gun10}. Thus
formula \rref{class} is perhaps to be interpreted by saying that one
has to use local sections to $f$ to transfer the standard hyperbolic
metric to $\Omega$ (as in ref. \cite{ZT} for the case of Schottky
uniformization) and then locally pull--back this last one down to $X.$

Bearing in mind these warnings, let us briefly mention how
various classical formulas \cite{S} arise in this formalism.
Consider the (improved) Liouville energy-momentum tensor:
\[
{\cal T}_\alpha(\za)=
(\partial\varphi_\alpha)^2-\partial^2\varphi_\alpha\0
\]
It is holomorphic on--shell. Plugging in \rref{class}, we
find:
\[
{\cal T}_\alpha(\za)=-\frac{1}{2}\left\{A_\alpha(\za),\za\right\}\0
\]
so that we can identify it with the projective connection $-\frac{1}{2}u_\alpha
(\za)$. Moreover, using the spin $-1/2$ realization of the projective atlas
$A_\alpha=\xi_{1\alpha}/\xi_{2\alpha}$, we can write the Liouville field as:
\be
e^{-\varphi_\alpha}=\xi_{1\alpha}\bar\xi_{2\alpha}-
\xi_{2\alpha}\bar\xi_{1\alpha}
\label{Liouvfield}
\ee
with
\[
\xi_{1\alpha}=\frac{A_\alpha}{(\partial A_\alpha)^{1/2}}\,\,\,\,,\,\,\,\,
\xi_{2\alpha}=\frac{1}{(\partial A_\alpha)^{1/2}}\0
\]

\subsection{A comparison between the two types of solutions}

Thus far in this subsection we have presented the uniformization point of
view concerning the Liouville equation. One might even superficially
conclude that the reconstruction formula \rref{Liouvfield} is the same
as the one in the previous subsection \rref{lsol}. Although the similarity
between the two formulas are certainly not accidental, there
are two important differences. The first one is that our approach in
the previous subsection is inspired by conformal field theory. As is
very profitably done in such a theory we split our problem into two {\it
independent} holomorphic and anti--holomorphic parts. In the language of this
section this amounts to considering the solutions of the Liouville
equation of the form
\[
e^{2\varphi_\alpha}=\frac{\partial A_\alpha(\za)\bar\partial
B_\alpha(\bar\za)}{(A_\alpha(\za) - B_\alpha(\bar\za))^2}
\]
where $B$ is independent of $A$.
It is obvious that in order to obtain a (1,1) form from this formula, one has
to put constraints on the monodromy of $\{B_\alpha\}$.
If one in addition requires reality of
$e^{2\varphi_\al}dz_\al d\bar z_\al$, the flat cocycles are again reduced
to $PSU(1,1)$
with the additional freedom of taking $B_\alpha(\bar z_\al)$ to be a
projective transform of $\overline{A_\alpha(\za)}$.

The second important difference with the uniformization setting is that we
actually consider only the solutions ensuing from the holomorphic and
anti--holomorphic DS linear systems. This allows us to represent the
solutions in terms of bosonic oscillators. We proved in \cite{ABBP}
in genus 0 that there is a one--to--one correspondence between the space of
free bosonic oscillators and hyperbolic solutions of the Liouville equation.
The same proof does not work in higher genus. A simple example will
show the difficulties we run into (on this point, see \cite{A}).

In order to extend the proof of \cite{ABBP}, one should be able
to `transform' the connection $\Lambda$ in \rref{lambda} (for simplicity
we drop the label $\al$) into the upper triangular connection typical
of the DS system. In this way one would end up with the bosonization
formula
\[
\xi_1=e^{-\wp}\,\,\,\,,\,\,\,\,\bar\partial\wp=0
\]
and would find the relation:
\be
-\partial^2\wp+(\partial\wp)^2=-\frac{1}{2}\{A,z\}={\cal T}(z)
\label{miura}
\ee
which is a {\em Miura transformation}. It is matter of coordinate patching
to prove that in order for \rref{miura} to be consistent, $\partial\wp$ should
transform as a holomorphic connection on $K^{-{1\over 2}}$ and, as we proved
above, such objects do not exist. Thus,
paying attention to the residues, $\partial\wp$ can be taken to be
meromorphic with simple poles. This is still not enough, since in a
coordinate transformation $\xi_1$ and $\xi_2$ mix with each other
and this is compatible with \rref{miura} if
and only if the transition functions $ \{T_{\alpha\beta}\}$
of the flat bundle $T$ are triangular
matrices. However this is hardly acceptable if $T$ has to be a flat bundle
arising from (or yielding) uniformization. For, in this case the
projective structure should actually be an affine one, which cannot
exist for a surface of genus $g>1$ \cite{gun1}.

The last remark, based on a monodromy argument, points toward a clear--cut
separation between the uniformization solution for a compact
Riemann surface and the solutions based
on the DS systems (which we rely on for quantizing the theory):
we are not going to find the uniformization solution among the latter.
In fact DS systems only define  (branched) affine structures on $X$ with
some points removed (see below), rather than projective structures (as the
uniformizing solutions do). In the next subsection we clarify this point.

\subsection{The DS system and branched affine structures}

An affine structure on a general surface $X$ is a collection of local
charts $\{ A_\al\}$ related by the transformation rule
\be
A_\alpha(\za)=a_{\alpha\beta}A_\beta(\zb)+b_{\alpha\beta}
\label{affine}
\ee
The target space is assumed to be $\CC ,$ although also $\PP^1$ can be
considered. An affine structure can be obviously considered as a
special kind of projective structure. However, as proved in
\cite{gun10,gun1}, affine structures do not exist whenever the surface
is compact (and of genus $g>1$), as their existence is equivalent to
the triviality of the canonical bundle $K.$ Allowing for branch
points, that is points where the local charts $\{ A_\al\}$ may not be
local homeomorphism, avoids that obstruction, so that these slightly
generalized structures do exist even on compact surfaces
\cite{mandel}.

{}From the analysis developed in section 2 and 3, it follows
that the DS system is related to affine structures rather than to
projective ones. Indeed, given a solution
$({\sigma_1}_\al ,{\sigma_2}_\al )$ of the DS system we can form the
maps
\be
f_\alpha = -\frac{{\sigma_2}_\al}{{\sigma_1}_\al}
\label{aff:map}
\ee
which from the monodromy properties previously analyzed have exactly
the transformation property \rref{affine}. Now we have
\[
\del f_\alpha = {\sigma_1}_\al^{-2}
\]
and since
\[
{\sigma_1}_\al^{-2}(Q)=
h_\alpha (Q)^2 h_\alpha (Q_\al )^{-2}
\exp\left(-2 p_n \int_{Q_\al}^Q \omega^n\right)
\]
we see that branch points occur at the zeroes of $h.$ Thus branch
points unavoidably appear, even if we consider a non closed
surface. The maps $\{ f_\alpha\}$ will in general be badly behaved in the
vicinity of the points $P_+$ and $P_-$ where the fields have essential
singularities. Note, though, that if we set to zero all the modes
$p_k$ except those corresponding to the holomorphic differentials in
the KN basis, we can undelete the points $P_+$ and $P_-$ and the only
singular points left will be the zeroes of $h.$ This is the standard
case treated in the mathematical literature (see \cite{mandel}).

The maps in \rref{aff:map} globalize
to give a (branched) developing map \cite{gun20}
\[
f:\tilde{X'}\longrightarrow\CC
\]
and the holonomy homomorphism into the affine group:
\[\rho :\pi_1(X')\longrightarrow A(1,\CC)\]
Actually, from our previous results we are
able to explicitly compute this homomorphism
\[\rho (\gamma )z=\cb^2_{\gamma^{-1}}z+\sb_{\gamma^{-1}}\]
for any $z\in\CC .$

\section*{}
\subsection*{Appendix A}

Here we explicitly calculate the cocycle $\sb_\gamma$. Before
doing this, it is necessary to see in some detail how to realize
the isomorphism
\[
H^1(M,\, G)\cong \mbox{\sl Hom}(\pi_1(M),\,G)/G
\]
The main point is to describe the fundamental group from within a
\v{C}ech setting. This is done in the following way. Fix a base-point
on the manifold and a suitable good-covering of which we fix a certain
element $U_0$ containing the base point. Given a closed path
$\gamma ,$ we cover it with a chain
$(U_0,U_1,\ldots ,U_n,U_0)$
of open sets starting and ending at $U_0.$ Homotopic paths correspond
to chains such that we can pass from one to the other by a finite
sequence of simple operations consisting in replacing a pair
$(U_i,\, U_{i+1})$ of consecutive elements by a triple
$(U_i,\, U_j,\, U_{i+1})$ with non void intersection. The inverse path
$\gamma^{-1}$ correspond to the chain
$(U_0,U_n,\ldots ,U_1,U_0),$ and so on. This defines the group
$\pi_1(\goth U , U_0).$ The true fundamental group is obtained by
taking the direct limit over the coverings $\goth U .$
Given the cocycle $g=\{ g_{ij}\}$ the corresponding element in
$\mbox{\sl Hom}(\pi_1(M),\,G)$ is the one assigning to the chain
$(U_0,U_1,\ldots ,U_n,U_0),$ the group element
$g_{01}\cdot g_{12}\cdots g_{n-1\, n}\cdot g_{n0}.$
It is clear that this operation is
well-defined up to conjugation and compatible with refinements of the
covering (full details in \cite{gun1}).

Next, in order to have the necessary formulas at hand, and also to put
the significance of ${\bf s}_{\gamma}$ in the right context, we
quote from \cite{gun2} the
following theorem: on any connected manifold $M$ carrying a flat
bundle $F,$ there is an isomorphism
\[
H^1(M ,{\sh F})\cong H^1(\pi_1(M),{\bf F})
\]
where $\sh F$ is the sheaf of locally constant sections of $F,$
${\bf F}$ is the associated characteristic representation and the
space at the RHS is the group of {\em crossed homomorphisms} of
$\pi_1(M)$ into the representation space of ${\bf F}$ \footnote{The
representation space of ${\bf F}$ can be taken
to be ${\sss\CC}^r$ where $r$ is the rank of $F.$}
modulo the trivial ones.

We recall that a crossed homomorphism $u$ of a group $\Pi$ into a
$\Pi$-module $V$ is a map $u:\Pi\rightarrow V$ satisfying
\[
u(xy)=y^{-1}\cdot u(x) +u(y)
\]
$x,\, y\in\Pi ,$ where the dot stands for the action of $\Pi$ on $V.$
The space of all crossed homomorphism is denoted by $Z^1(\Pi ,\, V).$
The trivial crossed homomorphisms (i.e. the coboundaries) are those
given by
\[
u(x)=v-x^{-1}\cdot v
\]
for $v\in V.$ The space of the coboundaries is denoted by
$B^1(\Pi ,\, V).$ Thus $H^1(\Pi ,\, V)$ $=$
$Z^1(\Pi ,\, V)/B^1(\Pi ,\, V).$

It is now worth describing the explicit form of the isomorphism
$H^1(M ,{\sh F})$ $\cong$ $H^1(\pi_1(M),{\bf F}).$ We use the
representation of $\pi_1(M)$ we are now familiar with, namely the one
given by $\pi_1({\goth U},U_0).$  For a cocycle
$\{ A_{\alpha\beta}\}\in Z^1({\goth U}, {\sh F}),$ the corresponding
element in $Z^1(\pi_1({\goth U},U_0), \hat{F})$ is given by
\bean
\hat{A}_\gamma &=& (\hat{F}_{\alpha_1\alpha_2}\cdots
\hat{F}_{\alpha_{p-1}\alpha_p})^{-1}\cdot A_{\alpha_0\alpha_1}+
(\hat{F}_{\alpha_2\alpha_3}\cdots \hat{F}_{\alpha_{p-1}\alpha_p})^{-1}
\cdot A_{\alpha_1\alpha_2}+\cdots\\
& &\cdots + \hat{F}_{\alpha_{p-1}\alpha_p}^{-1}\cdot
A_{\alpha_{p-2}\alpha_{p-1}}+A_{\alpha_{p-1}\alpha_p}
\eean
where $\gamma$ is the chain
$(U_{\alpha_0},U_{\alpha_1},\ldots ,U_{\alpha_p})$ based at $U_0.$
It is a standard matter of tracing all the relations to verify that
this correspondence is well defined on the classes and goes through
when taking the direct limit on the coverings. It is also easy to
verify the cocycle condition starting directly from the formula above.
We refer to \cite{gun2} for the details.

The point is now simply to insert the KN parametrization into the
expressions for $c_{\alpha\beta}$ and $d_{\alpha\beta}$ and apply the
general formulas we quoted above. To keep all the matter conceptually
(if not practically) more manageable, it is better to recast
$s=s_{\alpha\beta}$ in slightly different form,
perhaps giving it a nicer geometrical significance.
Indeed notice that the transformation rule \rref{glue1} for
$\s_1$ can be rewritten in the following form
\[
e^{-2\L_\alpha (\za )}d\za = c_{\alpha\beta}^{-2}
e^{-2\L_\beta (\zb )}d\zb
\]
and can be interpreted as the definition for each $\alpha ,\,\beta$ of
a 1-form $\phi_{\alpha\beta}:$
\[
\phi_{\alpha\beta}=
\left\{ \brr{ll}e^{-2\L_\alpha}d\za &\mbox{on $U_\alpha$}\\
c_{\alpha\beta}^{-2}e^{-2\L_\beta}d\zb &\mbox{on $U_\beta$}\err\right.
\]
It is easy to see that the two prolongations $\phi_{\alpha\beta},$
$\phi_{\alpha\gamma}$
from $U_\alpha$ agree on the intersection
$U_\beta\cap U_\gamma ,$ so that we can drop
the second index: $\phi_{\alpha\beta}\rightarrow\phi_\alpha .$
Inserting the explicit expression for $d_{\alpha\beta}$
and the definition of $s_{\al\beta},$ we find
\[
s_{\alpha\beta}= \int_{(Q_\beta ,Q_\alpha )}\phi_\beta
=-c_{\alpha\beta}^2\int_{(Q_\alpha ,Q_\beta )}\phi_\alpha
\]
where $(Q_\alpha ,Q_\beta )$ is the 1-simplex joining $Q_\alpha$
with $Q_\beta .$

At this point formula \rref{sgamma} is inferred
plugging the
equation above into the expression of ${\bf s}_\gamma$ as an element
of $Z^1(\pi_1({\goth U},U_0), {\bf c}^{-2}).$
Here are the first few steps.
Consider for instance three open sets $U_\alpha ,U_\beta ,U_\gamma .$
According to the quoted prescription we must consider
\[
c_{\beta\gamma}^2s_{\alpha\beta}+s_{\beta\gamma}
\]
which, using the previously introduced forms $\{\phi_\alpha\}$, reads
\bean
-c_{\beta\gamma}^2\int_{(Q_\alpha ,Q_\beta )}\phi_\beta
-\int_{(Q_\beta ,Q_\gamma )}\phi_\gamma
&=&-c_{\beta\gamma}^2\int_{(Q_\alpha ,Q_\beta )}\phi_\beta
-c_{\beta\gamma}^2\int_{(Q_\beta ,Q_\gamma )}\phi_\beta \\
&=&-c_{\beta\gamma}^2\int_{(Q_\alpha ,Q_\gamma )}\phi_\beta
\eean
so that, inserting the KN parametrization into $\phi_\beta ,$ we find
\[
h_\gamma (Q_\gamma )^{-2}e^{2\sum_n p_n \int_{Q_\alpha}^{Q_\gamma}\omega_n}
\int_{(Q_\alpha ,Q_\gamma )}
e^{-2\sum_k p_k \int_{Q_\alpha}^{z}\omega_k}h(z)^2
\]
Next add $U_\delta$ at the end of the chain. The relevant quantity now
is
\[
c_{\gamma\delta}^2c_{\beta\gamma}^2s_{\alpha\beta}
+c_{\gamma\delta}^2s_{\beta\gamma}+s_{\gamma\delta}
\]
and, using the result for three sets, we reexpress it as
\[
-c_{\gamma\delta}^2c_{\beta\gamma}^2\int_{(Q_\alpha ,Q_\gamma )}\phi_\beta
-c_{\gamma\delta}^2\int_{(Q_\gamma ,Q_\delta )}\phi_\gamma
\]
but the forms $\phi_\gamma$ and $c_{\beta\gamma}^2\phi_\beta$ are the
same on $U_\beta\cap U_\gamma$, so that the last expression can be
written in term of the integral of a unique form we keep calling
$\phi_\gamma :$
\[
c_{\gamma\delta}^2c_{\beta\gamma}^2s_{\alpha\beta}
+c_{\gamma\delta}^2s_{\beta\gamma}+s_{\gamma\delta}
=-c_{\gamma\delta}^2\int_{(Q_\alpha ,Q_\delta )}\phi_\gamma
\]
Plugging in again the expression in terms of the KN basis we find the
same formula as the one relative to the three sets, except for the
shift of indices:
\[
h_\delta (Q_\delta )^{-2}e^{2\sum_n p_n \int_{Q_\alpha}^{Q_\delta}\omega_n}
\int_{(Q_\alpha ,Q_\delta )}
e^{-2\sum_k p_k \int_{Q_\alpha}^{z}\omega_k}h(z)^2
\]
... and so on. It is clear that we obtain the formula \rref{sgamma}.

\subsection*{Appendix B}

Here we show that conditions \rref{condgamma}, \rref{Fgamma},
\rref{barFgamma} are {\em equivalent} to the univalence of the
solution obtained from \rref{lsol}. Sufficiency being obvious, only
necessity is to be proved. Thus assume that a collection
$\{ M_\alpha\}$ such that
\[
e^{-\varphi_\alpha}=
k_{\alpha\beta}^{-1/2}{\bar k}_{\alpha\beta}^{-1/2}
e^{-\varphi_\beta}
\]
indeed exists, where each $e^{-\varphi_\alpha}$ is given by
\rref{lsol}. This means that the quadratic form defined by \rref{lsol}
must be an $\Slc$--scalar, which in turn is true if and only if
\[
M_\beta = {}^tT_{\alpha\beta}\, M_\alpha\, {\bar T}_{\alpha\beta}.
\]
In the representation space
$\mbox{\sl Hom}(\pi_1(X'),\Slc )/\Slc$ this is reformulated by
saying that the characteristic representations $\bar {\bf T}$ and
${\bf T}^\vee$ must be equivalent, that is we must have
\[
M\, {\bar {\bf T}}_\gamma ={\bf T}^\vee_\gamma\, M
\]
for a certain $M\in\Slc$ and any $\gamma\in\pi_1(X').$ Writing down
the components of $M$ as
\[
M=\left(\brr{cc}x&y\\ u&v\err\right),\quad\quad det\,M=1
\]
and working out the matrix products, one easily derives the following
equations
\ban
\bar \cb_\gamma &=& \cb_\gamma^{-1}\\
\sb_\gamma &=& {y\over v}-{y\over v}\cb_\gamma^2\\
\bar \sb_\gamma &=& {u\over v}-{u\over v}\bar\cb_\gamma^2
\ean
A glance at Appendix A will immediately convince the reader that these
relations
exactly mean that the cocycles $\sb$ and $\bar\sb$ are coboundaries.
This proves necessity.

Now it is an immediate consequence that the characteristic
representations ${\bf T}^\vee$ and $\bar {\bf T}$ must be
diagonalizable. Indeed if $f={y\over v} ,$ and from the above coboundary
relations, it follows at once that
\[
{\bf T}_\gamma=
\left(\brr{cc}\cb_\gamma & 0\\
\cb_\gamma^{-1}\sb_\gamma & \cb_\gamma^{-1}\err\right)=
\left(\brr{cc}1&0\\ -f&1\err\right)\cdot
\left(\brr{cl}\cb_\gamma & 0\\ 0 & \cb_\gamma^{-1}\err\right)\cdot
\left(\brr{cc}1&0\\ f & 1\err\right)
\]
A similar relation holds for $\bar {\bf T}_\gamma $ with
${\bar f}={u\over v} .$ Diagonalizability means that the extension
classes represented by ${\bf T}^\vee$ and $\bar {\bf T}$ are trivial,
and so are the flat bundles $T$ and ${\bar T}.$

\subsection*{Appendix C}

In our analysis of the univalence for the Liouville solutions
in Appendix B and the Toda solutions in section 7, we understood
the hypothesis that the intertwiner $M$ is Gauss factorizable.
We also remarked how this condition is essential for the Bloch--wave
basis representation to exist. However, it is interesting to see what
happens if we drop this requirement. Let us analyze the Liouville case
in some detail.

Let $M$ be the matrix introduced in Appendix B.
The condition $v\neq 0$ is necessary and sufficient for $M$ to admit a
Gauss factorization in the form
\[
M=N_+\,D\, N_-
\]
with $D$ diagonal and $N_\pm$ upper and lower unipotent matrices.

Thus let us suppose that $v=0,$ so that
\[
M=\left(
\brr{cc}x&y\\
	u&0\err
\right)\, ,\qquad u=-\frac{1}{y}
\]
Then the familiar equation
\[
M\, {\bar {\bf T}}_\gamma ={\bf T}^\vee_\gamma\, M
\, ,\qquad \gamma\in\pi_1(X')
\]
gives the conditions:
\ba
\bar\cb_\gamma &=& \cb_\gamma\label{ns1}\\
\bar\sb_\gamma &=& \frac{1}{y^2}\sb_\gamma +
		    \frac{x}{y} (1-\cb_\gamma^2)\label{ns2}
\ea
Thus we see that in this case the conditions so obtained are such that
the antichiral monodromy must represent the same flat bundle as the
chiral one. Indeed, condition \rref{ns1} forces the diagonal elements
to be the same, while condition \rref{ns2} tells us that the
representations ${\bf T},$ ${\bar {\bf T}}$ differ by a conjugation in
the lower Borel subgroup of $\Slc .$ One has directly
\[
{\bar {\bf T}}_\gamma =
\left(
\brr{cl}y&0\\
	x&y^{-1}\err
\right)\cdot {\bf T}_\gamma \cdot
\left(
\brr{cc}y^{-1}&0\\
	-x    &y\err
\right)
\]
as desired. As for the flat bundles, this means that the extension
\ban
0\longrightarrow C^{-1}\longrightarrow T \longrightarrow
&C&\longrightarrow 0\\
0\longrightarrow C^{-1}\longrightarrow {\bar T} \longrightarrow
&C&\longrightarrow 0
\ean
differ by an automorphism.

It is to be noticed that in this case the flat bundle $T$ is allowed
to be a non trivial extension class.

Finally, it is worth mentioning the following similarity
with the uniformization formulas. In the special case $x=0,$ $y=-1$ we
have $\bar\sb_\gamma = \sb_\gamma$ so that $\bar T = T$ and
\[
e^{-\varphi_\alpha}=\sigma_{2\alpha}\bar\sigma_{1\alpha}-
\sigma_{1\alpha}\bar\sigma_{2\alpha}
\]

The class of solutions analyzed in this Appendix do {\it not} belong to the
phase space $\F_0$. For this reason we call them non--standard. The relevance
of these solutions for quantization is an open question.

\subsection*{Appendix D}

It is interesting to examine the conformal properties of the $\psi$--basis
in terms of the symplectic structure introduced in section 5.
Let us consider the energy momentum tensor
\[
{\cal T}= \pg^2 +\pg'
\]
which naturally appears in the DS system through the equation
\[
\partial^2 \s_i={\cal T}\s_i,\quad\quad\quad i=1,2
\]
As it turns out, $2{\cal T}$ is a projective connection.
According to the decomposition \rref{pg}, we can rewrite
\be
{\cal T}= p^2+\nabla p +{R_0\over 2},\quad\quad R_0=2(\Gamma_0' + \Gamma^2_0)
\label{R}
\ee
$R_0$ is a fixed projective connection, and $\nabla= \del +2n\Gamma_0$
represents in this Appendix the covariant derivative applied to the
weight $n$ tensors.
Since $p^2$ and $\nabla p$ are holomorphic two forms on $X'$, we can
expand them on the basis of quadratic differentials
\[
{\textstyle {1\over 2}} (p^2+\nabla p)= \sum_k \ell_k \Omega^k\0
\]
i.e.
\be
\ell_k = {\textstyle{1\over 2}}
 \sum_{n,m} l_k^{nm}p_np_m +{\textstyle{1\over 2}}\zeta_k^np_n\0
\ee
where
\[
l_k^{nm}= \pai\oint_{\ct} e_k\omega^n\omega^m, \quad\quad
\zeta_k^n =\pai \oint_{\ct}e_k\nabla \omega^n
\]
Using the basic Poisson brackets \rref{sympl} we find -- in this Appendix
we abandon the simplified notation for the Poisson brackets adopted in
section 5, therefore the following Poisson brackets are the usual ones --
\be
\{\ell_n, \ell_m\}= - C_{nm}^k \ell_k - {1\over {8\pi i}} \oint_{\ct}
\tilde\chi_0 (\ell_n,\ell_m)\label{KNl}
\ee
where $\tilde \chi_0$ has the same form as in eq.\rref{chi2} provided the
role of projective connection be played by $R_0$, eq.\rref{R}.
Therefore, up to an irrelevant -- sign, eq.\rref{KNl} represents a
realization of the extended KN algebra \rref{extKN}.

We can now easily work out, for example, the Poisson bracket
\be
\{\ell_n, \psi_1(Q)\}= e_n(Q) p(Q) \psi_1(Q)- {\textstyle{1\over 2}}
\nabla e_n(Q)\psi_1(Q)
\label{lpsi}
\ee
The RHS of this equation is nothing but $\L_{e_n}$, i.e. the Lie derivative
along the vector field $e_n$, applied to the weight $-{\textstyle{1\over 2}}$
tensor
$\psi_1(Q)$. This is exactly what we expect.

As for $\psi_2(Q)$ the calculation
is longer, but once the formalism is established we can safely rely on the
genus zero results \cite{BBT} which guarantee that $\psi_2(Q)$ behaves like a
weight $-{\textstyle{1\over 2}}$ tensor as well.
We stress that the $\s$--basis does
not have good tensorial properties with respect to the symplectic structure
introduced in section 5.

We remark that the above
equations \rref{KNl} and \rref{lpsi}, remain unchanged if we replace the
Poisson brackets with the Dirac brackets defined in subsection 5.5.
Indeed the Poisson brackets of the $\ell_n$'s with $\F_r$ and $\bar \F_r$
turn out to be proportional to $\F_r$ and $\bar \F_r$, respectively.
Thus the energy--momentum tensor is the generator of the conformal
transformations also with the correct bracket on the constrained
manifold in the phase space.

\end{document}